\documentclass[12pt,preprint]{aastex}

\usepackage{graphicx}
\usepackage{amsmath}
\input{epsf.sty}                        
\input{psfig.sty}                       
\shorttitle{Structure and Kinematics of the 43 GHz SiO Maser
around the VX Sgr} \shortauthors{Xi Chen et al.}

\begin{document}


\title {INWARD MOTIONS OF THE COMPACT SiO MASERS AROUND VX SAGITTARII}
\author
  {Xi ~Chen\altaffilmark{1, 2}, Zhi-Qiang ~Shen\altaffilmark{1}, Hiroshi ~Imai
  \altaffilmark{3}, Ryuichi ~Kamohara \altaffilmark{3,4}}

  \altaffiltext{1} {Shanghai Astronomical Observatory, 80 Nandan Road, Shanghai 200030, P.R.China}
  \altaffiltext{2} {The Graduate School of Chinese Academy of Sciences, Beijing 100034, P.R.China}
  \altaffiltext{3} {Department of Physics, Faculty of Science, Kagoshima University, 1-21-35 Korimoto, Kagoshima 890-0065, Japan}
  \altaffiltext{4} {VERA Project Office, National Astronomical Observatory, 2-21-1 Osawa, Mitaka, Tokyo 181-8588}

\label{firstpage}


\begin{abstract}

We report Very Long Baseline Array (VLBA) observations of 43 GHz
$v$=1, $J$=1--0 SiO masers in the circumstellar envelope of the
M-type semi-regular variable star VX Sgr at 3 epochs during 1999
April--May. These high-resolution VLBA images reveal a persistent
ringlike distribution of SiO masers with a projected radius of
$\approx$3 stellar radii. The typical angular size of 0.5 mas for
individual maser feature was estimated from two-point correlation
function analysis for maser spots. We found that the apparent size
scale of maser features was distinctly smaller than that observed
in the previous observations by comparing their fractions of total
power imaged. This change in the size scale of maser emission may
be related to stellar activity that caused a large SiO flare
during our observations. Our observations confirmed the asymmetric
distribution of maser emission, but the overall morphology has
changed significantly with the majority of masers clustering to
the north-east of the star compared to that lying to the
south-west direction in 1992. By identifying 42 matched maser
features appearing in all the three epochs, we determined the
contraction of an SiO maser shell toward VX Sgr at a proper motion
of $-0.507\pm$0.069 mas yr$^{-1}$, corresponding to a velocity of
about 4 km s$^{-1}$ at a distance of 1.7 kpc to VX Sgr. Such a
velocity is on the order of the sound speed, and can be easily
explained by the gravitational infall of material from the
circumstellar dust shell.

\end{abstract}


\section{Introduction}

Circumstellar SiO masers associated with late-type stars provide a
unique probe of the kinematics in the extended atmosphere.
Dominated by the mass-loss process and permeated by shocks,
magnetic fields and density gradients, the extended atmosphere is
a complex region located between the photosphere and the inner
dust formation shell. VLBI experiments have demonstrated that SiO
masers lie in ringlike configurations, presumably centered on the
star, with tangential rather than radial maser gain paths (Diamond
et al. 1994; Greenhill et al. 1995; Boboltz, Diamond \& Kemball
1997; Phillips et al. 2003; Soria-Ruiz et al. 2004; Boboltz \&
Wittkowski 2005). The maser ring is not static, its diameter
changes with phase of stellar pulsation, which has been observed
in Mira variables R Aqr (Boboltz et al. 1997) and TX Cam (Diamond
\& Kemball 2003).

VX Sagittarii (VX Sgr) is a semi-regular variable with an optical
period of 732 days (Kukarkin et al. 1970) and a spectral
classification between M4e-Ia and M9.8. The optical light curve of
VX Sgr exhibited some interesting quiescent phases intercepting a
regular 732-d oscillation during the past 75 years: 1945-1950,
1963-1967 and 1997-2002 (Kamohara et al. 2005). From
high-resolution optical interferometry data, Monnier et al. (2004)
fitted a stellar radius (R$_{\ast}$) of 4.35 mas, 3 times smaller
than that of 13 mas obtained by Greenhill et al. (1995). Its
circumstellar envelope exhibits strong OH, H$_{2}$O and SiO maser
radiation (Chapman \& Chohen, 1986; Greenhill et al. 1995;
Murakawa et al. 2003; Kamohara et al. 2005). The distance to VX
Sgr remains uncertain; for comparison with previous work by
Greenhill et al. (1995), we adopt a distance of 1.7 kpc, at which
VX Sgr can be classified as a red supergiant. The systemic
velocity of VX Sgr is estimated to be about 5.3 km s$^{-1}$ based
on a dynamical model of an expanding envelope fitting to the
interferometric maps of OH maser emission at 1612 MHz (Chapman \&
Cohen 1986).

The typical lifetime of an SiO maser feature, which is both
spatially and velocity coherent, is of order one month. This
limits the time interval of the monitoring observations to study
the kinematics of the extended atmosphere. In this paper, we
report 3-epoch (in 39 days) VLBI monitoring observations of the
SiO maser emission toward VX Sgr in 1999. The observations and
data reduction are described in $\S$ 2; results and discussion are
presented in $\S$ 3, followed by conclusions in $\S$ 4.

\section{Observations and data reduction}
The 43 GHz $v$=1, $J$=1--0 SiO maser emission toward VX Sgr was
observed at 3 epochs (1999 April 24, May 23 and May 31) using the
Very Long Baseline Array (VLBA) of the NRAO\footnote[1]{The
National Radio Astronomy Observatory is a facility of the National
Science Foundation operated under cooperative agreement by
Associated Universities, Inc.}. The data were recorded in left
circular polarization in an 8 MHz band and correlated with the FX
correlator in Socorro, New Mexico. The correlator output data had
256 spectral channels, corresponding to a velocity resolution of
0.22 km s$^{-1}$. A rest line frequency of 43.122027 GHz was
adopted for the maser transition.

The data reduction was performed using the Astronomical Image
Processing System (AIPS) package. The continuum calibrators were
used to calibrate phase gradients along frequency (group delays)
and bandpass characteristics. The amplitude calibration was done
by estimating the scaling factors necessary to best fit a time
series of the total power spectra for each station to a single
calibrated total power spectrum obtained from the Mauna Kea
station. A zenith opacity of about 0.05 at Mauna Kea was estimated
from the variation in system temperature with zenith angle during
these observations. Residual fringe-rates were obtained by
performing a fringe-fitting on a reference channel (at
V$_{LSR}$=0.1 km s$^{-1}$), which has a relatively simple
structure in the maser emission. At this stage in the reduction,
fringes are detected on all baselines of the VLBA including the
longest SC-MK baseline. In practice, several maser components were
overlapped in the reference channel. Therefore, self-calibration
on the reference channel was performed to remove any structural
phase. The solutions of fringe-fitting and self-calibration were
then applied to the data in all the velocity channels.

Synthesis images were constructed using the standard imaging
techniques. Image cubes consisting of individual channel maps were
made for velocities between 22 and $-$8 km s$^{-1}$. Uniformly
weighted visibility data created a synthesized beam of 0.18 mas
$\times$ 0.65 mas with a position angle $-9^\circ$. The detection
limit was 150 mJy beam$^{-1}$, which is about 5 times the noise
level in these images. The positions and flux densities of the
detected maser components in each channel map were estimated by
fitting a two-dimension Gaussian brightness distribution using the
AIPS task SAD. The typical uncertainty of the fitted position of
maser components was smaller than 10 $\mu$as.

\section{Results and Discussion}
\subsection{Maser spots and features}
\subsubsection{statistics for SiO maser spots and
features}

In each of the velocity channel maps, there are some spatially
distinguished components, each of which has a bright peak. The
component is called a ``maser spot''. We usually see several maser
spots that are clustered within a small region in space and
Doppler velocity, typically 1 AU and 1 km s$^{-1}$, respectively.
We call such a group of maser spots a ``maser feature''. Thus,
spots within the same velocity channel can belong to different
features. In order to study the characteristics of SiO masers,
individual features were determined for each of the three epochs.
The maser spots in different channels were deemed as the same
feature according to the following criterion, these spots (i)
appear in at least three adjacent channels, (ii) have peak flux
densities greater than 200 mJy (7$\sigma$), and (iii) lie within
0.5 mas (see Section 3.1.2). Finally, 110, 110, and 105 maser
features were selected for epochs 1, 2 and 3, respectively.

Table 1 lists both the average and median values of the sizes for
the identified maser spots and features and the velocity ranges
across features. The spot size was estimated from arithmetic
average of sizes of the major and minor axes of the spot image,
which were obtained by fitting to an elliptical Gaussian
brightness distribution in the CLEAN map. Figure 1 shows the
histograms of the spot size. A typical apparent size of a spot is
0.5 mas, corresponding to 0.85 AU at 1.7 kpc distance, which is
slightly larger than the geometric mean of the VLBA beam of 0.35
mas at 43 GHz. The size of a feature, $L$, is defined by
\begin{equation}
L=\sqrt{\theta^{2}_{s-s}+\sigma^{2}_f} \  ,
\end{equation}
where $\theta_{s-s}$ is the fitted size of the brightest spot in
the feature, and $\sigma_f$ is the standard deviation in the
estimated feature positions. Thus, the feature size indicates the
extent of the distribution of the accompanying spots. It can be
seen from Table 1 that the feature sizes are almost the same as
the spot sizes, of $\sim$0.5 mas, suggesting significant
clustering of spots in the feature. The velocity range across the
feature, $\Delta u$, is defined to be the difference between the
maximal and minimal velocities of the spots in the feature. A
typical value of $\Delta u$ is 0.86-1.74 km s$^{-1}$ (see column
(4) in Tables 2, 3 and 4), corresponding to 4-8 spots in an
individual feature, with the average and median values of 1.4-1.6
and 1.3-1.5 km s$^{-1}$ (see Table 1).

The full lists of parameters measured for each feature are
available in Tables 2, 3 and 4, for epochs 1, 2 and 3,
respectively. We fit a Gaussian curve to the velocity profile of
feature containing at least four spots to determine V$_{LSR}$ at
the peak of velocity profile and its uncertainty
$\sigma_{V_{LSR}}$, full width at half maximum (FWHM) $\Delta$V
and its uncertainty $\sigma_{\Delta V}$ and peak flux density $P$.
For some features which can not be well represented by a Gaussian
curve (labelled by a ``$\ast$'' in Tables 2, 3 and 4), intensity
weighted mean V$_{LSR}$ and uncertainty $\sigma_{V_{LSR}}$=0.03 km
s$^{-1}$ were adopted; the FWHM velocity $\Delta$V was equal to
$\Delta u$/2, with uncertainty $\sigma_{\Delta
V}$=68\%$\Delta$V=34\%$\Delta u$; the flux density of the
brightest spot in each feature was deemed as its peak flux
density. Feature positions (x, y) in right ascension (R.A.) and
declination (Dec.) were determined from an intensity weighted
average over maser spots in the feature. The uncertainty
($\sigma_{x}, \sigma_{y}$) of a feature position was defined as
squared root of the square sum of (1) the mean spot distance from
the defined feature position and (2) the mean measurement error of
the spot positions. The weights proportional to the intensity of
the spot were applied in the uncertainty estimation. The typical
position uncertainties of features are 0.01 mas and 0.02 mas for
R.A. and Dec., respectively.  The positions are measured with
respect to the reference feature at (0, 0) (labelled by a ``r'' in
Tables 2, 3, and 4).

\subsubsection{Two-Point Correlation Function of Maser
Spots}

To determine the criterion of angular separation for identifying
the maser features, we performed the ``two-point correlation
function'' analysis. The technique has been used in the analysis
of distributions of water masers in W49N (Walker 1984; Gwinn
1994), and W3 IRS5 (Imai, Deguchi \& Sasao 2002). The two-point
correlation function of maser spots describes the number of spots
per unit angular area (1$''$$\times$1$''$) with a given
separation, $\Delta r=(\Delta x^{2}+\Delta y^{2})^{1/2}$, on the
sky (x-y) plane from an arbitrary spot. This function can be
expressed as (Imai, Deguchi \& Sasao, 2002)

\begin{equation}
n_{s}(\Delta r)d\Omega=\frac{\sum_{i,j}n_{\delta}(\mid
\overrightarrow{r_{i}}-\overrightarrow{r_{j}}\mid)}{n_{spot}}
\end{equation}
where
\begin{equation}n_{\delta}(r)=
\begin{cases} 1 & \text{when}\ \Delta r<r<\Delta r+dr, \\0 & \text{otherwise}.
\end{cases}
\end{equation}
Here dr is a separation of $\Delta r$ between the successive bins
in the $n_{s}(\Delta r)$ plot, $d\Omega=2\pi\Delta rdr$ is the
area of annulus, $n_{spot}$ is the total spot number, and the
indices $i, j$ run over all spots.

Figure 2 shows the two-point correlation functions of SiO maser
spots around VX Sgr at each epoch. The function for spots can be
fitted by a power law, $n_{s}(\Delta r)=n_{0}\Delta r^{\alpha}$,
on certain ranges. Interestingly, we found two power-law fits on
scale ranges of 0.03-0.25 mas and 0.5-20 mas at each epoch,
corresponding to a linear-scale range of 0.05-0.43 AU and 0.85-34
AU. The two steep slopes with an index $\alpha\approx-1.1\sim-1.2$
show that the maser spots are strongly clustered within the
corresponding scales. The distribution between 0.25 and 0.5 mas is
consistent with being flat, indicating no clustering on these
scales. The break of the power-law at 0.25 mas suggests a
characteristic scale size for clustering of spots to make a
feature. Therefore, the angular diameter should be 0.5 mas for
each feature. Consequently, combining spots separated by less than
0.5 mas yields physical maser features. Another large region of
0.5-20 mas is more likely to be the clusters of features since it
covers almost the entire mapping region of the detected masers.
Because the uncertainty of relative spot positions is 0.01 mas,
the plot shown in Figure 2 at separations less than 0.01 mas may
not be accurate.

From above two-point correlation function analysis, we obtained a
typical angular diameter of 0.5 mas for clustering of spots to
create maser features. But, this criterion may be uncertain for
some maser features. In addition, this technique does not include
any information of Doppler velocity of masers. Thus it only
provides a crude criterion for identifying maser features.
Actually, in the procedure of identifying maser features, we also
examined the intensity-velocity profiles of features and, we
treated them as separate features if they were distinctly multiple
peaks within 0.5 mas.

\subsection{Images of the SiO maser Shell}
Figure 3 shows the distributions of maser features for three
epochs. A persistent ringlike distribution of SiO masers around VX
Sgr with the strongest emission lying to the north-east (NE) of
the star can be seen from these images. The SiO maser shell should
be close to, but above, the stellar photosphere of VX Sgr. If we
adopt a stellar radius R$_{\ast}$=4.35 mas and an inner radius of
dust shell of 60 mas (Monnier et al. 2004), an estimated radius of
12.5 mas (Fig. 3) indicates that the maser shell lies close to the
stellar surface within about 3 R$_{\ast}$ and of course in the
inner dust shell. This is well consistent with the typical 2-4
R$_{\ast}$ for late-type stars (Diamond et al. 1994).

In the upper panel of Figure 4, we compare the total power imaged
by the VLBA (filled circle) to the total power (solid line)
obtained from the Mauna Kea antenna. The total power imaged is
obtained by fitting 2-dimension Gaussian components (maser spots)
in all the channels and then by summing all spots belonging to
features for a given channel rather than flux collected from image
pixels. Lower panel shows that the fraction of total power imaged
(solid line) is mostly between 0.5 and 0.9. On average, about
$\sim$70\% of the total luminosity of masers were detected in our
observations. As shown in Figure 1, the apparent sizes of maser
spots mostly ranged from 0.3 to 0.7 mas, with a mean of 0.5 mas.
This mean scale size is slightly larger than the geometric mean of
the VLBA beam of 0.35 mas at 43 GHz. Therefore, $\sim$30\% missing
flux in our map is mainly due to the high spatial resolution of
the interferometric array. That is, most of maser components are
detected in our VLBI observations and have a compact scale size of
about 0.5 mas. Greenhill et al. (1995) detected some dense
velocity coherent structures with a characteristic size of
$\sim$0.2 mas in the extended atmosphere; but 74\% of the maser
flux was resolved by their observations of the 43 GHz SiO $v$=1
$J$=1--0 transition in 1992, suggesting that the extended
atmosphere also had a lot of undetected masers with the angular
scales larger than about 3 mas. Doeleman et al. (1998) showed that
only $\sim20\%$ of the total flux density were detected in a
single baseline VLBI observation of the 86 GHz SiO $v$=1 $J$=2-1
transition in 1994, inferring that most of 86 GHz SiO maser
emission were larger than the beam size of 0.4 mas. Because the
flux density threshold is about 1.8 Jy (5$\sigma$) on each channel
image in the observations of Greenhill et al. (1995), for
comparison, we also calculate the fraction of total power imaged
with the 1.8 Jy threshold for 2-dimension Gaussian components
(open circle and dotted line in Figure 4). We find that the
fraction of total power imaged for components with the 1.8 Jy
threshold (dotted line) is almost the same as that for all maser
components (solid line). Hence, our result is significantly
different from that of Greenhill et al. (1995), implying that the
apparent size scale of maser features became distinctly smaller in
our observations than the previous ones. It should be noted that
the earlier VLBI observations suffered from some kind of
electronic noise in the VLBI systems that decorrelated the
signals, which would result in a net loss in the observed flux
density.

Kamohara et al. (2005) showed that the 43 GHz $v$=1, $J$=1--0 SiO
masers flared around 1999 March, which was within the 6-yr
(1997-2003) quiescent phase when the regular pulsation was
terminated. They proposed that the flare of SiO masers was due to
the increase of the gas ejected from the photosphere to excite SiO
maser emission under collisional pumping in a shock. If this model
is correct, it can be applied to the diminished size of maser
emission toward VX Sgr observed in 1999. The larger kinetic energy
of the outflow from the photosphere may produce larger turbulent
motions in maser regions in 1999. Such turbulent motions can then
create steeper velocity gradients in a smaller coherent path
length in a postshock region, i.e., forming the compact maser
features. This can also explain why more extended maser clouds
were seen from VLBI observations in 1992 August (Greenhill et al.
1995), because a weak outflow can be inferred from the observed
low SiO maser emission presumably due to the decrease in the mass
loss rate (Kamohara et al. 2005).

We also notice that the SiO maser emission in the NE is distinctly
stronger than that in the south-west (SW). This suggests an
asymmetric mass loss rate, i.e., the mass loss toward the NE
should be more significant in our observation sessions. This is
quite different from the VLBI observations in 1992 (Greenhill et
al. 1995) and 1994 (Doeleman et al. 1998), which showed major
maser emission concentrating in the SW region and thus suggested a
larger mass loss from the SW part. Combined together, these data
show a remarkable change in the major mass loss direction from the
SW (in 1992 and 1994) to the NW in 1999. However, the mechanism
responsible for the variation of the mass loss from the SW to the
NE remains to be understood. This may also be related to the
optical stellar phase and can be clarified in the future with
multi-epoch VLBI observations.

\subsection{Proper Motion}
From Figure 3, it is apparent that some SiO maser features live
longer than 39 days, the maximum time separation among three
epochs. This enables us to study their proper motions and the
kinematics of the circumstellar envelope of VX Sgr by comparing
the matched features that appear in all the three epochs.

Because of the nature of standard VLBI data reduction, the
absolute position of the phase center in each image is lost. For
studying the proper motion, we must align multi-epoch maps. The
feature used for registration is the one with a velocity
V$_{LSR}\approx$0 km s$^{-1}$ (labelled by a ``r'' in Tables 2, 3
and 4) at all the three epochs. This feature was chosen because of
its similarity in both morphology and velocity. The coordinate
frames for the three epochs were then shifted to align the origin
(0, 0) with this feature. At an assumed distance of 1.7 kpc, a
proper motion of 0.1 mas in 39 days corresponds to 7.5 km
s$^{-1}$. The expansion velocity of the H$_{2}$O masers, estimated
from a standard expanding outflow model, is 10 and 20 km s$^{-1}$
at the inner and outer of the H$_{2}$O maser shell, respectively
(Murakawa et al. 2003). The maximum proper motion for SiO masers
should be less than these values according to the standard
expanding outflow model (Chapman \& Cohen 1986). Thus the
allowable change in position is less than 0.13 mas during our
observations, corresponding to the largest possible velocity of 10
km s$^{-1}$. Allowing the maximum position uncertainty of (0.05,
0.07) mas (see Tables 2, 3 and 4), the angular separation during
all observation sessions should not exceed 0.2 mas. We can match
these features from one epoch to another using this criterion.
However, there may be some maser features lying within the angular
separation of less than 0.2 mas at a single epoch, this gives some
difficulties to match them among three epochs. Thus, we also
examined the velocity profiles and flux densities of the likely
matched features. We only selected the matched feature with the
similarity in both velocity profile and flux density if a feature
at one epoch has multiple possible matched features at another. As
a result, 58 commonly matched maser features were found between
first two epoches and 42 features were identified among all three
epoches (see Tables 2, 3 and 4). This suggests that lifetime of
SiO maser features around VX Sgr could be larger than one month.
Assuming an exponential decay of SiO maser features, we estimated
a scale lifetime of 44$\pm$4 days from our 3-epoch data.

Actually, the process of aligning maps for all the three epochs on
a feature located on the shell could introduce a systematic bias
in the proper motion estimate; individual maser proper motions are
uncertain by a constant offset vector that represents the motion
of the reference feature. If we assume that the average of motions
for all matched features represents the motion of the reference
feature, it was only about (-0.01, 0.01) mas in (R.A., Dec.)
during our observation sessions (39 days), and can be ignored
compared to the possible position errors discussed above. For the
following proper motion study, 42 matched maser features detected
in all three observing sessions were chosen. Figure 5 shows the
velocity profiles of these 42 SiO maser features. The velocity
profile and flux density for each matched feature is very similar
from one epoch to another.

One way to look into the global proper motion is to compute
separations between pairwise combinations of features. The
pairwise separations technique has been applied to SiO masers
toward R Aqr (Boboltz et al. 1997) and has no dependence on the
alignment of maps. In this procedure, a separation between two
features at one epoch and that between the corresponding features
at another epoch are calculated. The difference between these two
separations is referred to as the pairwise separation. When the
calculation is repeated for all the possible combinations of
matched feature pairs, mean values of these pairwise separations
are $-0.013\pm$0.003, $-0.017\pm0.003$ and $-0.030\pm$0.003 mas in
an interval of 8, 31 and 39 days, respectively, where the
uncertainties are the standard error of the mean. By performing a
weighted linear least-squares fit to these values, we obtained a
proper motion of $-0.171\pm$0.048 mas yr$^{-1}$. However, the
fitted motion may decrease toward zero due to the bias caused by
calculating all the possible pairs, including small separations.
To minimize this effect and determine a representative value for
the shift due to the proper motion, we only computed pairwise
separations for those pairs separated by more than 12 mas, here 12
mas represents the radius of maser shell. The separation of 12 mas
is important because the centers of the pairs are located on the
star and the pairwise separations of such pairs directly indicates
the global motion of the SiO maser shell. Figures 6 (a), (b) and
(c) show results from those pairs separated by more than 12 mas
over time intervals of 8, 31 and 39 days, respectively. As
indicated by bold lines in Figure 6, the mean of each of these
distributions is negative, implying an overall contraction of the
maser shell. These mean shift values are $-0.015\pm$0.004,
$-0.045\pm$0.004 and $-0.060\pm$0.005 mas in an interval of 8, 31
and 39 days, respectively. This gives a proper motion
$-0.507\pm$0.069 mas yr$^{-1}$ or a velocity of $-4.1\pm$0.6 km
s$^{-1}$. Actually, this velocity should be taken as the upper
limit to the true contraction because the pairwise separations for
those pairs separated by more than the radius of maser shell would
be greater than the real separation of the maser ring. In above
analysis of proper motion, the uncertainty quoted for the mean
shift of pair separations is the standard error, which is
different from the method used by Boboltz et al. (1997). For
comparison, we obtained an infall velocity of 4.2$\pm$0.2 km
s$^{-1}$ for those pairs separated by more than 12 mas, by using
exactly the same method of Boboltz et al. (1997). These are
consistent with our estimates with even smaller errors. Thus, as a
conservative estimation, we adopt the standard error in the paper.

We conclude that the maser shell contracts toward the star with a
velocity of about 4 km s$^{-1}$. However, converting from proper
motion $\mu$ (mas yr$^{-1})$ to velocity $v$ (km s$^{-1}$) depends
on the distance D (kpc), $v=4.74\mu\cdot D$. Under the assumption
of a distance of 1.7 kpc, the infall velocity of 4 km s$^{-1}$ in
VX Sgr is similar to 4.2$\pm$0.9 km s$^{-1}$ in R Aqr (Boboltz et
al. 1997). This speed of contraction can be easily obtained
through the gravitational infall. For the central star mass of 10
M$_{\odot}$ (Chapman \& Cohen 1986), the velocity of a particle
gravitationally falling from the inner dust shell at 60 mas to the
SiO maser shell at 12 mas would be about 25 km s$^{-1}$. Assuming
a temperature of 1500 K (Doel et al. 1995) and adiabatic condition
of SiO maser region, the sound speed should not exceed 3 km
s$^{-1}$. Thus, the inward proper motion of SiO maser shell is on
the order of the sound speed. In addition, our measured infall
velocity is comparable to the shock velocity in the radio
photospheres of long-period variables of $<$ 5 km s$^{-1}$ at a
distances beyond approximately 2 R$_{\ast}$ (Reid \& Menten 1997).

The possible existence of a bipolar outflow was speculated by
Kamohara et al. (2005) to explain the distribution and the
velocity structure of SiO masers toward VX Sgr seen at a single
epoch. After adding more data overlapping their observing epoch,
we have measured an inward proper motion of SiO masers toward VX
Sgr, which excludes the proposed bipolar outflow model. Actually,
the kinematics in SiO maser regions could be very complicated. An
SiO maser movie of Mira variable TX Cam has displayed the
co-existence of both local infall and outflow motions at a certain
stellar phase (Dimond \& Kemball 2003). Our current data are
insufficient to reveal such a complicated kinematics in VX Sgr.

The contraction of SiO maser shell has been reported in two Mira
variables R Aqr (Boboltz et al. 1997) and TX Cam (Diamond \&
Kemball 2003). Our observations provide the first direct evidence
for an inward motion of SiO maser shell around red supergiant
variable star. The corresponding optical pulsation phases of three
VLBA observations that revealed an inward motion toward R Aqr are
$\phi$=0.78, 0.87 and 0.04, respectively (Boboltz et al. 1997).
Diamond and Kemball (2003) found the expansion of SiO maser shell
in TX Cam with a ballistic gravitational deceleration over the
stellar phase interval from $\phi$=0.7 to 1.5; but beyond the
optical minimum at $\phi$ $\sim$1.5, a new shell forms interior to
the previously expanding outer shell. For the motion of the maser
shell around VX Sgr in 1999, the accurate stellar phase is unknown
because of the temporary quench of the optical pulsation from 1997
to 2003. Assuming that the pulsating period was kept even in the
quiescent phase, we estimated the stellar optical phase of our
observation sessions to be from 0.75 to 0.80 (Kamohara et al.
2005). Thus, we can see that the stellar optical phase of red
supergiant VX Sgr is very close to that of Mira variables R Aqr
and TX Cam when the SiO maser shell contracts. These results can
be compared with that seen in the theoretical kinematical model
adopted in Humphreys et al. (2002). Our analysis shall provide
another test for such a theoretical model.

In order to investigate the effect of stellar pulsation on the SiO
maser emission, Humphreys et al. (1996, 2002) and Gray \&
Humphreys et al. (2000) coupled an SiO maser model (Doel et al.
1995; Gray et al. 1995) to an M-type Mira hydrodynamic pulsation
model (Bowen 1988). These numerical simulations show that masing
material starts to contract after a shock wave in the envelope
arrives at the maser zone, and such an infall under gravity lasts
for about one third of the stellar pulsation cycle; while for the
remainder of the cycle, a shock-driven expansion of the maser ring
dominates. So the contraction of the SiO maser shell only appears
in a certain stellar phase range, during which a shock front
disrupts the existing maser ring and new features then form in the
postshock region. The newly forming maser ring has a smaller
angular extent than previously existing ring. Humphreys et al.
(2002) quote a mean stellar phase of the arrival of the shock in
the SiO maser zone of $\phi=0.71\pm0.15$. However, these
simulations have a difficulty in relating the model phase to
optical stellar phase. That is, these simulations did not offer
the accurate optical stellar phase range when the SiO maser shell
contracts. The relationship between the optical and model phases
could be reliably fixed by a combination of interferometric
observations and theory in the future. For the moment, the
analysis of the inward proper motions of SiO masers toward VX Sgr,
R Aqr and TX Cam suggests that the contraction of SiO maser shell
could occur during an optical stellar phase $\phi$=0.5 to 1.

\section{Conclusions}
We summarize the main results obtained from 3-epoch (in 39 days)
monitoring observations of the 43 GHz $v=1, J=1-0$ SiO
circumstellar maser emission toward VX Sgr, covering a stellar
optical phase range from 0.75 to 0.80.

\begin{enumerate}
\renewcommand{\theenumi}{(\arabic{enumi})}

\item
 The two-point correlation function of spots shows a power-law relation on scales of 0.03-0.25 mas and 0.5-20 mas.
The break of the power-law at 0.25 mas suggests an angular size of
0.5 mas for clustering of spots to make a feature.
  \item
  Our observations confirmed a persistent ringlike structure of SiO
masers with a projected radius of about 3 stellar radii
(R$_{\ast}$), which is consistent with the typical 2-4 R$_\ast$
for late-type stars.
  The overall morphology has changed significantly with the majority of masers appearing in North-East of the star in 1999, compared to that
  lying to the South-West direction in 1992 and 1994, suggesting the change of mass loss rate over $\sim$5-7
years.

  \item
  We found that the apparent size scale of maser features in 1999 is distinctly
  smaller than that observed in 1992, by comparing their fractions of total power imaged.
  This may be related to stellar activity that caused a large SiO flare during our observations.
  \item
 Analysis of pairwise separation of 42 matched features suggests that the maser shell contracts toward
  VX Sgr with a velocity of about 4 km s$^{-1}$. This infall velocity is similar in number to that seen in R
Aqr. The stellar optical phase of red supergiant VX Sgr is very
close to that of Mira variables R Aqr and TX Cam when the SiO
maser shell contracts.
\end{enumerate}

\vspace{6mm} We thank an anonymous referee for helpful comments
that improved the manuscript. Z.-Q. Shen also acknowledges the
support by the One-Hundred-Talent Program of Chinese Academy of
Sciences.

\newpage


\begin{figure*}
\resizebox{100mm}{100mm}{\includegraphics{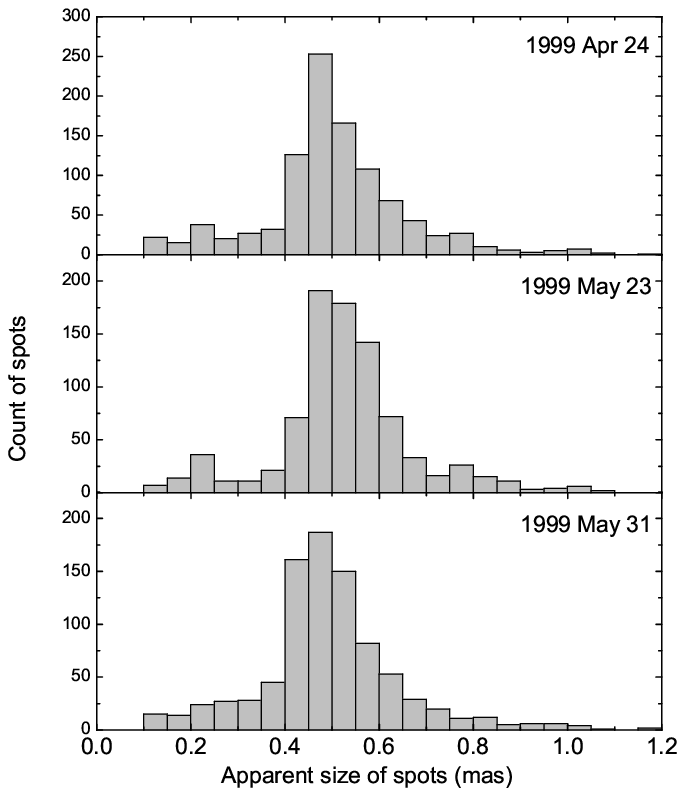}}
\caption{Histograms of SiO maser spot sizes in VX Sgr}
\end{figure*}

\newpage
\begin{figure}
\resizebox{110mm}{140mm}{\includegraphics {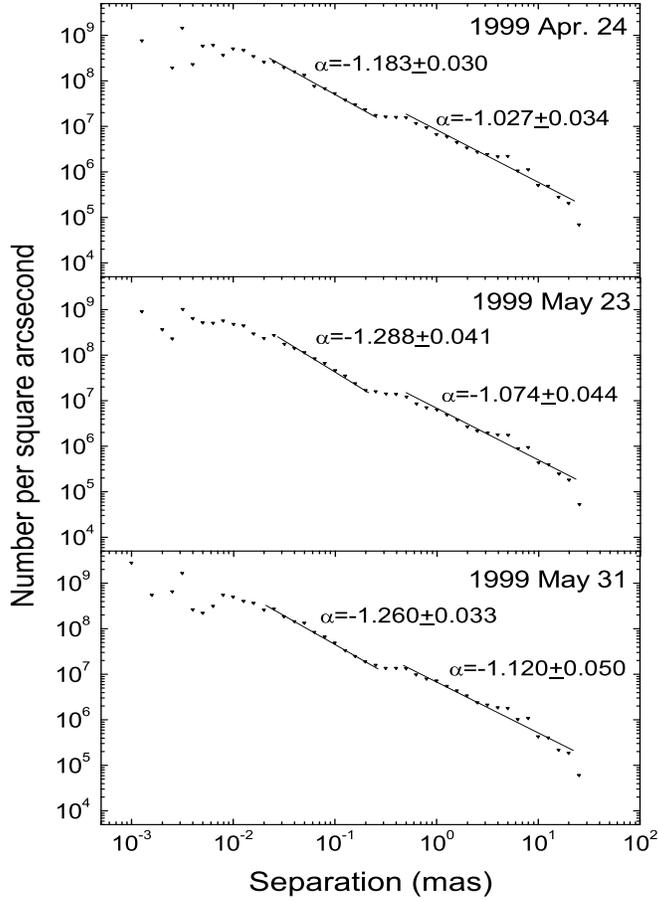}}

\caption{Two-point correlation functions for SiO maser spots in VX
Sgr. Solid lines show power law fits for spots in scale ranges of
0.03-0.2 mas and 0.5-20 mas, and the corresponding index $\alpha$
is also shown. The break of the power law at 0.25 mas indicates an
angular diameter of 0.5 mas for clustering of spots to make a
feature. The positional accuracy of the spots is 0.01 mas.}
\end{figure}

\begin{figure*}
\includegraphics[scale=0.95]{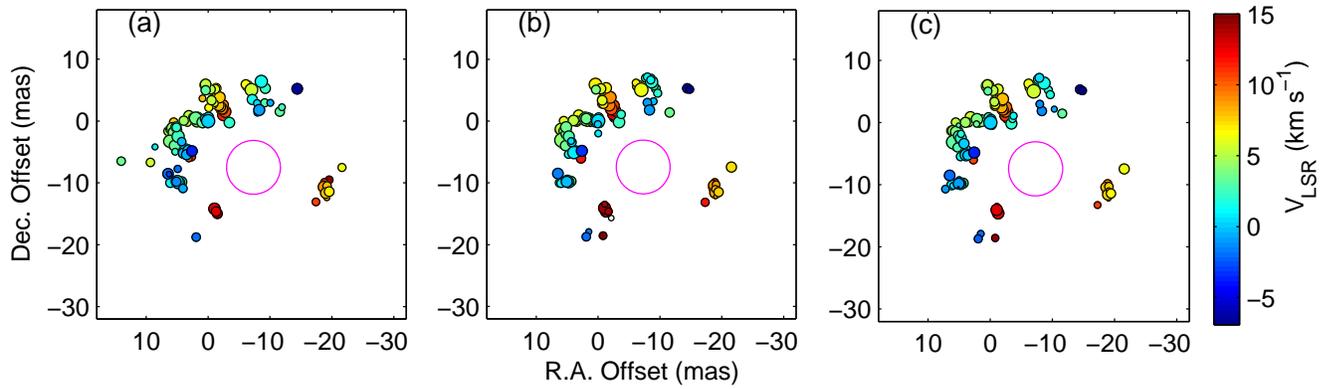}

\caption{VLBI images of 43 GHz {\it v}=1, {\it J}=1--0 SiO maser
emission toward VX Sgr obtained on (a) 1999 April 24, (b) 1999 May
23 and (c) 1999 May 31. Each maser feature is represented by a
filled circle whose area is proportional to its flux density in
logarithm, and the color indicates its Doppler velocity with
respect to the local standard of rest. Its stellar velocity is
about 5.3 km s$^{-1}$. The inner circle represents an observed
stellar disk of 8.7 mas in diameter (Monnier et al. 2004).}

\end{figure*}

\begin{figure*}
\resizebox{170mm}{90mm}{\includegraphics{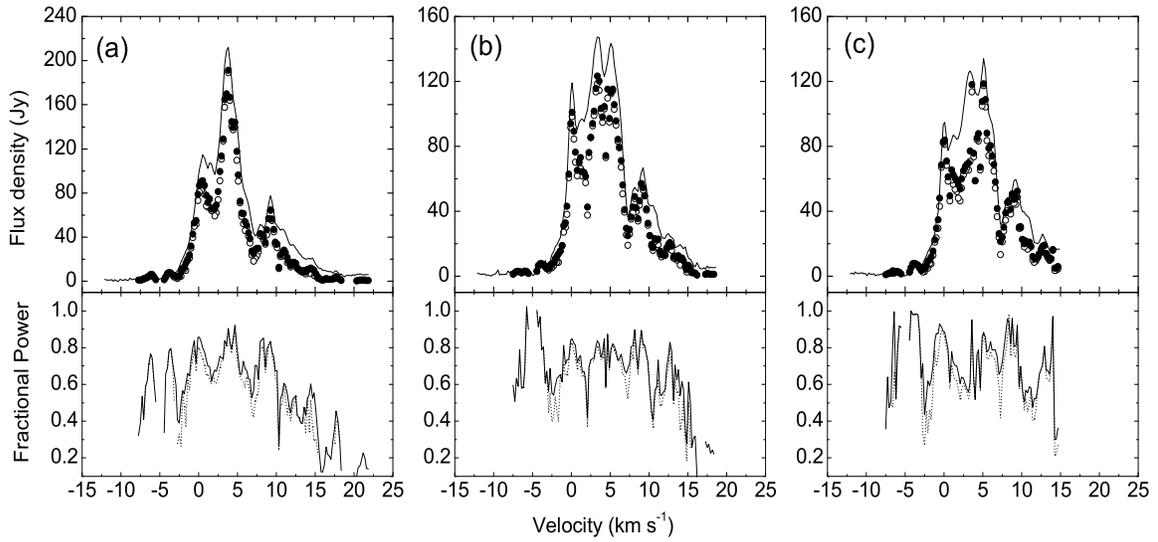}}

\caption{Upper panel: Comparison of total power (solid line) to
cross power (both open and filled circles) of 43 GHz {\it v}=1,
{\it J}=1--0 SiO maser emission toward VX Sgr obtained on (a) 1999
April 24, (b) 1999 May 23, and (c) 1999 May 31. Lower panel: The
corresponding fraction of total power detected by the
high-resolution VLBA observations. Filled circles (upper panel)
and solid line (lower panel) denote the cross power and the
fraction for all the detected maser features, respectively, while
open circles (upper panel) and dotted line (lower panel) represent
the corresponding cross power and fraction for the maser
components greater than 1.8 Jy in flux density.}
\end{figure*}


\begin{figure*}
\resizebox{100mm}{100mm}{\includegraphics{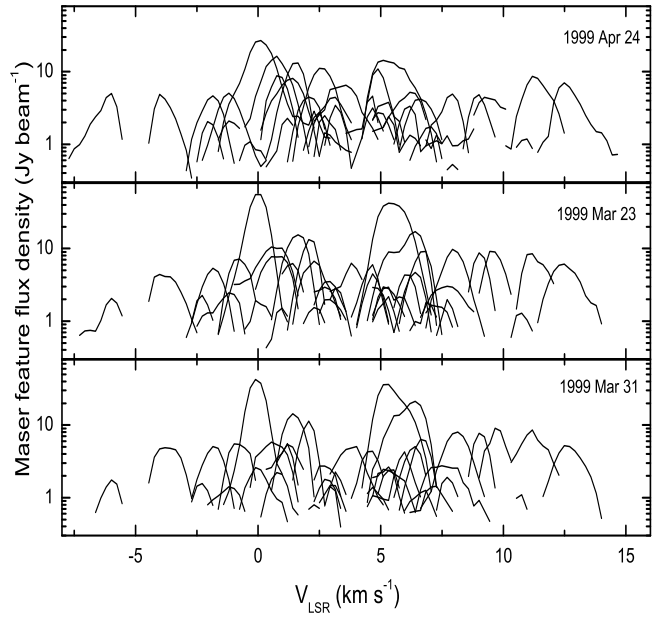}}

\caption{The velocity profiles of each matched maser feature
observed in three epochs.}
\end{figure*}

\begin{figure*}
\includegraphics[scale=1.3]{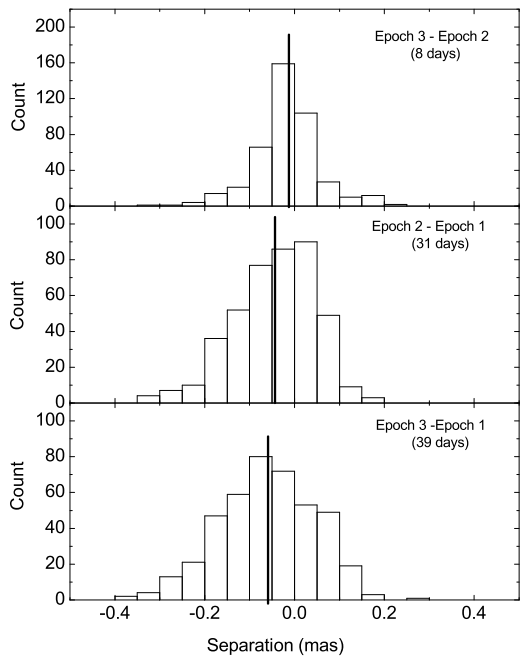}
\caption{Histograms showing the change in pairwise component
separations for those pairs separated by more than 12 mas, for (a)
Epoch 2 to Epoch 3, (b) Epoch 1 to Epoch 2, and (c) Epoch 1 to
Epoch 3. Bias toward negative shifts indicates a contraction of
the maser shell. Bold line represents the mean negative shift for
each of three histograms.}
\end{figure*}

\clearpage


\tabletypesize{\scriptsize}
\begin{deluxetable}{ccccccccccc@{}ccc}

 \setlength{\tabcolsep}{0.05in}
 \tablewidth{0pt}

\tablecaption{SiO maser spots and features in VX Sgr (a distance
of 1.7 kpc is assumed)} \startdata \hline\hline
 & & &\multicolumn{2}{c} {Spot Size}& & & &\multicolumn{2}{c}{Feature Size}& & &\multicolumn{2}{c}{Feature Velocity Range}\\
\cline{3-6} \cline{8-11} \cline{13-14}
  &  &  Average&  Average&  Median&  Median& &  Average&  Average&  Median &Median& & Average& Median\\
   Epoch& N$_{s}$$^{a}$&(mas)&(AU)&(mas)&(AU)& N$_{f}$$^{b}$& (mas)&(AU)&(mas)&(AU)&&(km s$^{-1}$)&(km s$^{-1}$)\\
 \hline
1 &   780&   0.51 &  0.87& 0.51 &  0.87&  110& 0.53&  0.90  & 0.50   & 0.85    &\  & 1.61    & 1.52\\
2  &   720 &  0.53 &  0.90  &0.52&  0.88 &  110& 0.55&0.94 & 0.52&   0.88  &\   & 1.42    & 1.33\\
3 &   670 & 0.50&  0.85 &  0.48&  0.82 &  105 & 0.50& 0.85 & 0.48&   0.82&\     &  1.41&    1.33\\

\enddata
\tablenotetext{a}{Number of maser spots}

\tablenotetext{b}{Number of maser features}

\end{deluxetable}

\clearpage


 \tabletypesize{\tiny}

\setlength{\tabcolsep}{0.06in}

\begin{deluxetable}{lrrrrrrrrrrrrrr}

\tablewidth{0pt} \tablecaption{\small 43 GHz SiO maser features
around VX Sgr observed by the VLBA on 1999 April 24.}

\tablehead{ID&\multicolumn{1}{r}{V$_{LSR}$}& $\sigma_{V_{LSR}}$
&$\Delta u$&$\Delta$V& $\sigma_{\Delta
V}$ &\multicolumn{1}{c}{x}&$\sigma_{x}$ &\multicolumn{1}{c}{y}&$\sigma_{y}$&\multicolumn{1}{c}{$L$}&P&S&\multicolumn{2}{c}{Match ID} \\
\cline {14-15}

&\multicolumn{2}{c}{(km s$^{-1}$)}&(km
s$^{-1}$)&\multicolumn{2}{c}{(km s$^{-1}$)}& \multicolumn{2}{c}{(mas)}&\multicolumn{2}{c}{(mas)}&(mas)&(Jy)&(Jy km s$^{-1}$)&Epoch 2&3\\
(1)&(2)&(3)&(4)&(5)&(6)&\multicolumn{1}{c}{(7)}&(8)&\multicolumn{1}{c}{(9)}&(10)&(11)&(12)&(13)&\multicolumn{2}{c}{(14)}
}

\startdata

      1&     21.14&     0.04&     1.74&     0.87&     0.27&    -6.819&     0.008&   -19.732&     0.021&      0.47&       1.3&       8.1&    &    \\
      2&     17.75&     0.01&     1.52&     0.55&     0.04&    -1.113&     0.004&   -15.783&     0.012&      0.46&       3.8&      14.5&      1&    \\
      3$^{*}$&     17.47&     0.03&     0.65&     0.33&     0.22&    -6.737&     0.010&   -24.075&     0.029&      0.48&       0.7&       2.1&    &    \\
      4&     16.63&     0.07&     1.09&     1.14&     0.78&    -0.676&     0.012&   -17.328&     0.026&      0.49&       0.7&       6.6&    &    \\
      5$^{*}$&     16.62&     0.03&     0.65&     0.33&     0.22&    -1.437&     0.008&   -14.512&     0.029&      0.45&       0.8&       2.1&    &    \\
      6&     14.91&     0.03&     2.17&     0.65&     0.07&    -0.822&     0.011&   -14.311&     0.011&      0.51&       5.3&      24.7&    &    \\
      7&     14.67&     0.03&     1.30&     0.54&     0.12&    -0.919&     0.008&   -18.247&     0.020&      0.55&       2.0&       9.0&      3&    \\
      8&     14.47&     0.02&     1.09&     0.63&     0.11&   -19.553&     0.004&    -9.521&     0.014&      0.47&       1.7&       6.3&    &    \\
      9&     14.16&     0.01&     1.52&     1.49&     0.69&    -1.483&     0.011&   -14.986&     0.015&      0.52&       7.8&      23.8&    &    \\
     10&     12.78&     0.05&     1.95&     1.55&     0.55&    -1.280&     0.003&   -14.632&     0.014&      0.48&       3.9&      21.6&    &    \\
     11&     12.61&     0.04&     3.48&     0.95&     0.11&    -1.010&     0.003&   -14.191&     0.009&      0.49&       9.3&      75.9&      8&      5\\
     12&     12.38&     0.02&     1.09&     0.71&     0.24&   -18.858&     0.010&   -10.848&     0.013&      0.54&       3.6&      13.2&    &    \\
     13&     11.41&     0.03&     2.61&     1.34&     0.21&    -2.476&     0.008&     1.001&     0.015&      0.55&      13.7&      91.1&     11&      7\\
     14&     11.03&     0.05&     0.87&     0.30&     0.20&   -18.777&     0.005&    -9.871&     0.015&      0.49&       1.9&       5.7&    &    \\
     15&     10.91&     0.02&     0.87&     0.41&     0.27&   -17.394&     0.006&   -13.077&     0.017&      0.50&       2.0&       6.5&     12&     10\\
     16&     10.87&     0.02&     1.09&     0.39&     0.09&     2.764&     0.025&    -5.737&     0.030&      0.57&       2.9&      11.8&    &    \\
     17&     10.33&     0.03&     1.96&     0.68&     0.12&    -2.752&     0.016&     1.517&     0.016&      0.80&      14.7&      57.8&     13&    \\
     18$^{*}$&      9.92&     0.03&     0.65&     0.33&     0.22&     3.126&     0.019&    -6.055&     0.021&      0.51&       1.8&       4.6&    &    \\
     19&      9.56&     0.01&     1.52&     0.41&     0.02&    -2.447&     0.014&     2.603&     0.028&      0.39&      15.0&      30.6&     16&     14\\
     20&      9.39&     0.02&     1.52&     0.55&     0.09&    -2.224&     0.011&     2.008&     0.022&      0.61&      31.7&     128.9&     17&    \\
     21&      9.00&     0.01&     1.52&     0.67&     0.04&    -0.816&     0.002&     3.522&     0.013&      0.52&       9.0&      35.9&     20&     18\\
     22&      8.61&     0.02&     1.74&     0.57&     0.07&    -1.948&     0.016&     2.327&     0.029&      0.59&      16.4&      56.2&    &    \\
     23&      8.02&     0.02&     1.52&     0.83&     0.10&    -1.442&     0.008&     4.007&     0.018&      0.63&       5.0&      24.3&     23&    \\
     24&      8.02&     0.05&     2.39&     0.68&     0.09&   -19.006&     0.008&   -11.645&     0.024&      0.50&      11.1&      42.2&     27&     26\\
     25&      7.99&     0.03&     1.52&     2.05&     4.02&   -18.705&     0.002&   -10.504&     0.007&      0.44&       6.2&      32.8&     24&    \\
     26&      7.99&     0.05&     0.87&     0.69&     0.24&    -2.224&     0.014&     1.989&     0.032&      0.45&      21.8&      56.1&    &    \\
     27$^{*}$&      7.96&     0.03&     1.09&     0.54&     0.37&   -19.172&     0.015&   -12.393&     0.024&      0.37&       1.0&       3.1&    &    \\
     28$^{*}$&      7.93&     0.03&     0.65&     0.33&     0.22&   -19.192&     0.036&   -10.487&     0.054&      1.43&       7.5&       9.7&     26&     28\\
     29&      7.86&     0.03&     1.96&     0.63&     0.10&    -1.874&     0.012&     3.840&     0.011&      0.51&       9.7&      47.2&     25&     25\\
     30&      7.86&     0.03&     1.74&     0.55&     0.08&    -1.681&     0.008&     2.554&     0.012&      0.54&       8.3&      33.7&     22&    \\
     31&      7.74&     0.01&     0.87&     0.40&     0.03&     0.904&     0.007&     3.643&     0.019&      0.48&       1.3&       4.4&    &    \\
     32$^{*}$&      7.70&     0.03&     1.09&     0.54&     0.37&     5.554&     0.017&    -0.181&     0.022&      0.57&       2.3&       9.7&    &    \\
     33$^{*}$&      7.17&     0.03&     0.87&     0.43&     0.30&    -0.591&     0.005&     2.829&     0.019&      0.52&       2.5&       7.8&     30&     30\\
     34&      6.84&     0.02&     1.52&     0.54&     0.09&   -19.553&     0.007&   -11.422&     0.024&      0.77&       3.6&      14.5&     31&     31\\
     35&      6.60&     0.04&     1.96&     0.91&     0.25&    -6.365&     0.004&     5.655&     0.005&      0.45&       6.2&      34.4&     32&     32\\
     36&      6.45&     0.04&     1.74&     0.83&     0.25&    -6.083&     0.004&     5.928&     0.011&      0.44&       3.6&      18.7&    &    \\
     37&      6.43&     0.02&     1.30&     0.62&     0.11&    -0.786&     0.010&     2.773&     0.017&      0.66&       5.5&      19.4&     34&    \\
     38$^{*}$&      6.42&     0.03&     0.65&     0.33&     0.22&    -0.114&     0.029&     2.157&     0.025&      0.53&       2.4&       7.5&    &    \\
     39&      6.40&     0.06&     1.30&     0.49&     0.33&     5.028&     0.008&    -1.440&     0.021&      0.61&       4.4&      15.7&    &    \\
     40&      6.28&     0.04&     1.09&     0.70&     0.43&   -21.632&     0.004&    -7.501&     0.016&      0.47&       2.0&       8.1&     35&     33\\
     41&      6.19&     0.03&     2.39&     0.84&     0.11&     0.431&     0.004&     5.829&     0.014&      0.48&       7.7&      43.9&     37&     36\\
     42&      6.11&     0.03&     1.30&     0.68&     0.25&    -1.270&     0.005&     5.285&     0.010&      0.44&       4.8&      19.2&     38&     38\\
     43$^{*}$&      5.99&     0.03&     0.65&     0.33&     0.22&    -1.106&     0.020&     5.189&     0.021&      0.81&       5.7&      15.6&    &    \\
     44$^{*}$&      5.98&     0.03&     0.65&     0.33&     0.22&    -1.072&     0.008&     2.873&     0.019&      0.50&       1.6&       4.2&    &    \\
     45$^{*}$&      5.61&     0.03&     0.87&     0.43&     0.30&     3.375&     0.008&    -0.204&     0.022&      0.63&       3.5&       8.5&     39&    \\
     46&      5.61&     0.03&     1.09&     0.58&     0.24&     9.343&     0.004&    -6.678&     0.012&      0.49&       2.4&       9.4&    &    \\
     47&      5.45&     0.04&     3.26&     1.42&     0.16&    -6.999&     0.003&     5.063&     0.005&      0.48&      22.0&     174.9&     40&     40\\
     48&      5.39&     0.08&     0.87&     0.70&     0.14&     0.412&     0.012&     5.933&     0.020&      0.50&       9.5&      33.3&    &    \\
     49$^{*}$&      5.37&     0.03&     0.87&     0.43&     0.30&    -0.410&     0.011&     5.122&     0.018&      0.61&       5.1&      16.1&     45&     43\\
     50&      5.32&     0.06&     1.52&     0.74&     0.40&     3.134&     0.007&     0.891&     0.012&      0.56&       8.5&      31.9&     42&     42\\
     51&      5.26&     0.05&     1.09&     1.45&     0.65&     6.405&     0.005&    -1.626&     0.022&      0.46&       6.5&      15.8&     53&    \\
     52&      5.26&     0.03&     0.87&     0.70&     0.47&     4.796&     0.007&    -2.225&     0.021&      0.52&       1.9&       6.8&    &    \\
     53&      4.94&     0.01&     1.09&     0.69&     0.08&    -0.547&     0.005&     3.340&     0.010&      0.48&       3.9&      16.5&     44&     41\\
     54&      4.92&     0.02&     2.39&     0.65&     0.07&     1.797&     0.004&    -0.044&     0.010&      0.53&      19.6&      60.3&     49&     51\\
     55&      4.67&     0.02&     1.09&     0.43&     0.11&     3.964&     0.004&     0.075&     0.007&      0.47&       4.7&      15.1&     52&     48\\
     56&      4.65&     0.02&     1.52&     0.75&     0.14&     5.790&     0.013&    -3.775&     0.012&      0.27&       4.4&      19.6&    &    \\
     57&      4.61&     0.01&     1.52&     2.61&     1.44&     5.625&     0.004&    -2.553&     0.021&      0.29&      15.5&      45.8&    &    \\
     58&      4.50&     0.05&     2.82&     1.00&     0.15&     5.661&     0.005&    -2.649&     0.014&      0.18&      10.9&      31.3&    &    \\
     59$^{*}$&      4.45&     0.03&     1.09&     0.54&     0.37&     6.025&     0.011&    -1.322&     0.031&      0.63&       8.8&      27.8&     53&     53\\
     60&      4.24&     0.02&     1.09&     0.33&     0.04&     2.370&     0.015&     0.584&     0.052&      0.48&       5.0&      16.8&    &    \\
     61&      4.10&     0.06&     1.74&     0.90&     0.51&     5.825&     0.015&    -1.132&     0.006&      0.47&       8.2&      44.4&    &    \\
     62&      3.93&     0.03&     3.69&     1.24&     0.09&     5.758&     0.004&    -3.184&     0.003&      0.52&      74.5&     550.5&     58&    \\
     63&      3.73&     0.03&     3.04&     1.13&     0.12&     5.465&     0.003&    -1.864&     0.009&      0.54&      19.9&     131.5&    &    \\
    64&      3.68&     0.02&     1.96&     0.57&     0.09&     6.222&     0.005&    -3.250&     0.012&      0.47&      28.2&     112.3&    &    \\
     65&      3.50&     0.05&     2.82&     0.96&     0.15&     2.026&     0.011&     0.476&     0.011&      0.63&      19.9&     114.2&     57&     56\\
     66&      3.47&     0.05&     3.26&     1.00&     0.16&     1.786&     0.011&     0.439&     0.009&      0.55&      35.6&     177.5&     62&    \\
     67&      3.34&     0.04&     1.96&     0.83&     0.16&     0.896&     0.008&    -0.185&     0.007&      0.51&      12.6&      61.9&     61&    \\
     68&      3.31&     0.07&     1.96&     1.34&     0.41&     5.400&     0.011&    -3.989&     0.022&      0.55&       8.0&      50.9&     68&     64\\
     69&      3.29&     0.06&     1.09&     1.44&     0.67&    14.045&     0.012&    -6.473&     0.014&      0.52&       2.6&      11.3&    &    \\
     70&      3.28&     0.05&     1.30&     0.97&     0.41&     0.406&     0.008&     5.000&     0.021&      0.48&       4.2&      17.3&     65&     60\\
     71&      3.17&     0.03&     1.52&     2.06&     0.62&   -11.558&     0.012&     1.507&     0.028&      0.56&       3.1&      17.5&     67&     61\\
     72&      3.05&     0.02&     0.87&     0.24&     0.16&    -9.136&     0.005&     2.954&     0.010&      0.45&       3.1&       9.1&    &    \\
     73&      2.93&     0.03&     1.30&     0.62&     0.14&     5.212&     0.005&    -0.963&     0.010&      0.47&       7.5&      27.7&     70&    \\
     74$^{*}$&      2.62&     0.03&     0.87&     0.43&     0.30&   -11.922&     0.008&     2.225&     0.019&      0.42&       1.3&       4.3&    &    \\
     75&      2.58&     0.03&     1.52&     0.42&     0.07&     0.384&     0.015&     0.453&     0.032&      0.69&       4.9&      21.7&    &    \\
     76&      2.54&     0.02&     2.61&     0.95&     0.08&     4.862&     0.006&    -2.501&     0.007&      0.50&      18.7&     116.6&     76&     62\\
     77&      2.45&     0.04&     2.17&     0.66&     0.10&     4.317&     0.010&    -9.857&     0.015&      0.58&       5.9&      33.3&     71&     66\\
     78$^{*}$&      2.39&     0.03&     0.87&     0.43&     0.30&     5.347&     0.010&    -4.956&     0.037&      0.36&       1.3&       4.5&    &    \\
     79&      2.31&     0.05&     1.30&     0.66&     0.38&    -9.234&     0.008&     5.303&     0.016&      0.55&       5.9&      24.7&     75&     71\\
     80&      2.06&     0.03&     1.09&     1.37&     0.30&     5.139&     0.007&    -1.037&     0.013&      0.44&       3.3&      12.1&    &    \\
     81&      2.06&     0.01&     1.74&     0.85&     0.06&    -3.417&     0.007&    -0.244&     0.008&      0.49&       7.3&      36.3&     77&     74\\
     82&      1.91&     0.03&     1.96&     0.85&     0.17&    -0.002&     0.007&     0.601&     0.021&      0.64&       7.2&      31.9&     74&    \\
     83&      1.54&     0.01&     1.30&     0.59&     0.02&    -7.121&     0.002&     3.493&     0.004&      0.42&       8.8&      29.4&    &    \\
     84&      1.46&     0.01&     2.39&     0.87&     0.06&    -8.600&     0.004&     6.395&     0.008&      0.50&      21.1&     107.3&     83&     78\\
     85&      1.38&     0.01&     2.39&     0.82&     0.11&     4.738&     0.007&    -9.977&     0.009&      0.48&      10.5&     100.1&     84&     79\\
     86&      1.12&     0.01&     1.30&     0.86&     0.06&     6.113&     0.018&    -9.991&     0.022&      0.55&       3.9&      17.0&     88&     84\\
     87$^{*}$&      1.03&     0.03&     0.87&     0.43&     0.30&     0.081&     0.015&    -0.666&     0.044&      0.38&       1.1&       3.0&    &    \\
     88&      0.90&     0.02&     1.74&     1.32&     0.65&     3.959&     0.008&    -5.067&     0.007&      0.45&      10.4&      49.6&     91&     82\\
     89&      0.84&     0.01&     0.87&     0.33&     0.15&     5.015&     0.023&   -10.357&     0.017&      0.21&       2.0&       4.4&    &    \\
     90&      0.74&     0.04&     2.82&     0.94&     0.14&     4.981&     0.002&    -9.771&     0.004&      0.45&      18.9&     126.6&     93&     88\\
     91$^{*}$&      0.74&     0.03&     0.65&     0.33&     0.22&     8.593&     0.005&    -4.193&     0.017&      0.46&       1.2&       3.1&    &    \\
     92$^{*}$&      0.72&     0.03&     0.65&     0.33&     0.22&    -7.969&     0.004&     1.370&     0.013&      0.47&       2.1&       5.3&    &    \\
     93&      0.47&     0.05&     1.52&     1.97&     0.82&     3.903&     0.007&    -4.423&     0.036&      0.30&       1.7&       8.2&    &    \\
     94&      0.46&     0.05&     2.17&     0.61&     0.13&     3.626&     0.010&    -4.652&     0.010&      0.68&      19.9&     103.5&     94&    \\
     95$^{*}$&      0.43&     0.03&     0.87&     0.43&     0.30&    -7.921&     0.009&     2.883&     0.027&      0.56&       1.6&       5.1&    &    \\
     96$^{r}$&      0.07&     0.01&     4.78&     0.96&     0.02&     0.000&     0.002&     0.000&     0.005&      0.48&      35.6&     243.2&     98&     91\\
     97&     -0.14&     0.02&     1.52&     0.61&     0.07&     4.013&     0.005&    -5.216&     0.030&      0.59&       4.6&      31.7&    &    \\
     98&     -0.33&     0.07&     1.30&     0.48&     0.23&     4.275&     0.009&    -3.324&     0.021&      0.50&       2.2&       9.6&     96&     89\\
     99&     -0.52&     0.03&     1.52&     0.57&     0.11&     3.509&     0.014&    -5.435&     0.024&      0.58&       7.2&      25.0&    &    \\
    100$^{*}$&     -0.54&     0.03&     0.65&     0.33&     0.22&   -10.051&     0.008&     2.962&     0.020&      0.54&       1.7&       5.0&    &    \\
    101&     -0.60&     0.03&     2.61&     0.83&     0.09&     3.197&     0.011&    -4.776&     0.012&      0.58&       9.4&      48.7&    &    \\
    102&     -0.70&     0.01&     1.09&     0.71&     0.02&     4.082&     0.002&   -10.888&     0.007&      0.43&       3.0&      11.0&    &    \\
    103&     -0.90&     0.03&     1.09&     2.11&     1.43&     5.114&     0.003&    -9.818&     0.009&      0.50&       3.7&      14.8&    102&     95\\
    104&     -1.22&     0.01&     2.82&     0.79&     0.02&    -8.239&     0.004&     1.805&     0.010&      0.49&       7.8&      45.9&    103&     98\\
    105&     -1.94&     0.02&     1.30&     1.23&     0.34&     1.945&     0.004&   -18.794&     0.011&      0.44&       2.4&      10.0&    107&    102\\
    106&     -2.00&     0.02&     2.17&     1.34&     0.34&     6.491&     0.004&    -8.510&     0.008&      0.47&       6.8&      35.2&    106&    100\\
    107$^{*}$&     -2.07&     0.03&     0.65&     0.33&     0.22&     4.910&     0.009&    -7.760&     0.012&      0.51&       1.9&       5.0&    &    \\
    108$^{*}$&     -2.94&     0.03&     0.65&     0.33&     0.22&     6.326&     0.005&    -8.597&     0.019&      0.46&       1.1&       2.9&    &    \\
    109&     -3.71&     0.03&     1.96&     0.94&     0.15&     2.612&     0.007&    -4.860&     0.010&      0.48&       8.3&      41.1&    108&    103\\
    110&     -6.17&     0.04&     2.39&     0.75&     0.10&   -14.401&     0.003&     5.235&     0.010&      0.45&       6.2&      35.5&    109&    104\\
\enddata

\tablecomments{column (1): ID number; columns (2) and (3):
V$_{LSR}$ at the peak of velocity profile and its uncertainty
$\sigma_{V_{LSR}}$; column (4): the velocity range across the
feature $\Delta u$; columns (5) and (6): the FWHM velocity
$\Delta$V and its uncertainty $\sigma_{\Delta V}$; columns (7) and
(9): the intensity weighted centroid of each feature ($x, y$);
columns (8) and (10): the corresponding uncertainties
($\sigma_{x}, \sigma_{y}$); column (11): the angular size of each
feature $L$; column (12): the peak flux density of each feature
$P$; column (13): the integrated flux density of all spots in the
feature S;  and column (14): the ID numbers of matched features at
other two epoches, determined by criteria in section 3.3.}

\tablenotetext{*}{Feature which can not be well represented by a
Gaussian curve.}

\tablenotetext{r} {Reference feature.}

\end{deluxetable}


\begin{deluxetable}{lrrrrrrrrrrrrrr}

\tablewidth{0pt} \tablecaption{\small 43 GHz SiO maser features
around VX Sgr observed by the VLBA on 1999 May 23.}

\tablehead{ID&\multicolumn{1}{r}{V$_{LSR}$}& $\sigma_{V_{LSR}}$
&$\Delta u$&$\Delta$V& $\sigma_{\Delta
V}$ &\multicolumn{1}{c}{x}&$\sigma_{x}$ &\multicolumn{1}{c}{y}&$\sigma_{y}$&\multicolumn{1}{c}{$L$}&P&S&\multicolumn{2}{c}{Match ID} \\
\cline {14-15}

&\multicolumn{2}{c}{(km s$^{-1}$)}&(km
s$^{-1}$)&\multicolumn{2}{c}{(km s$^{-1}$)}& \multicolumn{2}{c}{(mas)}&\multicolumn{2}{c}{(mas)}&(mas)&(Jy)&(Jy km s$^{-1}$)&Epoch 1&3\\
(1)&(2)&(3)&(4)&(5)&(6)&\multicolumn{1}{c}{(7)}&(8)&\multicolumn{1}{c}{(9)}&(10)&(11)&(12)&(13)&\multicolumn{2}{c}{(14)}
}\startdata

      1$^{*}$&     17.78&     0.03&     1.30&     0.65&     0.44&    -1.150&     0.005&   -15.676&     0.012&      0.49&       1.6&       7.8&      2&    \\
      2$^{*}$&     15.96&     0.03&     0.65&     0.33&     0.22&    -1.827&     0.009&   -15.747&     0.024&      0.48&       0.9&       2.5&    &    \\
      3&     14.85&     0.03&     1.09&     0.36&     0.08&    -0.929&     0.012&   -18.419&     0.029&      0.59&       1.3&       6.0&      8&    \\
      4$^{*}$&     14.00&     0.03&     0.65&     0.33&     0.22&    -2.117&     0.010&   -15.698&     0.028&      0.55&       1.1&       2.9&    &    \\
      5&     13.93&     0.01&     1.09&     0.64&     0.08&    -0.793&     0.011&   -18.550&     0.027&      0.55&       1.3&       6.1&    &    \\
      6&     13.59&     0.06&     3.48&     2.14&     0.55&    -1.292&     0.007&   -14.613&     0.015&      0.52&       6.9&      68.8&    &    \\
      7$^{*}$&     12.72&     0.03&     0.65&     0.33&     0.22&    -1.658&     0.012&   -14.664&     0.027&      0.61&       1.9&       5.2&    &    \\
      8&     12.60&     0.03&     4.13&     0.93&     0.09&    -1.014&     0.004&   -14.094&     0.011&      0.54&       8.9&      82.8&     11&      5\\
      9&     12.60&     0.06&     1.74&     1.20&     0.64&    -1.047&     0.005&   -13.455&     0.008&      0.24&       0.9&       3.2&    &    \\
     10&     12.46&     0.14&     1.30&     1.85&     0.98&    -2.620&     0.020&     0.597&     0.028&      0.59&       5.8&      19.8&    &    \\
     11&     11.36&     0.04&     1.74&     1.27&     0.99&    -2.437&     0.010&     1.101&     0.020&      0.51&      11.5&      67.5&     13&      7\\
     12&     10.83&     0.03&     1.09&     1.59&     0.50&   -17.282&     0.007&   -13.148&     0.023&      0.52&       2.1&       8.2&     15&     10\\
     13&     10.66&     0.03&     1.74&     0.96&     0.19&    -2.767&     0.013&     1.656&     0.025&      0.67&      10.0&      48.6&     17&    \\
     14&     10.61&     0.03&     1.52&     0.61&     0.11&     2.793&     0.021&    -6.060&     0.025&      0.53&       3.0&      16.8&    &    \\
     15&     10.28&     0.04&     1.74&     1.55&     0.92&   -18.665&     0.015&   -10.742&     0.047&      0.61&       5.9&      20.2&    &    \\
     16&      9.51&     0.02&     1.52&     0.72&     0.12&    -2.483&     0.011&     2.471&     0.018&      0.53&      14.1&      76.6&     19&     14\\
     17&      9.34&     0.04&     2.17&     0.90&     0.15&    -2.102&     0.018&     1.967&     0.013&      0.56&      17.5&      88.4&     20&    \\
     18$^{*}$&      9.24&     0.03&     0.65&     0.33&     0.22&   -18.862&     0.006&    -9.939&     0.020&      0.60&       2.5&       6.9&    &    \\
     19&      9.22&     0.01&     0.87&     0.61&     0.14&   -18.715&     0.019&   -11.049&     0.028&      0.62&       3.8&      13.1&    &    \\
     20&      9.08&     0.01&     1.30&     0.61&     0.01&    -0.834&     0.002&     3.511&     0.008&      0.51&      15.2&      53.8&     21&     18\\
     21&      8.37&     0.07&     1.74&     0.82&     0.35&    -1.894&     0.006&     3.247&     0.018&      0.33&       2.3&       8.6&    &    \\
     22&      8.33&     0.02&     1.52&     0.95&     0.07&    -1.739&     0.008&     2.491&     0.009&      0.51&      14.1&      58.7&     30&    \\
     23&      8.22&     0.05&     1.52&     0.46&     0.14&    -1.512&     0.013&     3.924&     0.019&      0.64&       5.1&      20.8&     23&    \\
     24&      8.03&     0.02&     1.74&     0.95&     0.21&   -18.650&     0.015&   -10.414&     0.010&      0.51&       5.0&      27.2&     25&    \\
     25&      8.00&     0.02&     2.61&     0.95&     0.08&    -1.947&     0.006&     3.832&     0.007&      0.49&      13.0&      80.3&     29&     25\\
     26$^{*}$&      7.79&     0.03&     1.09&     0.54&     0.37&   -19.068&     0.005&   -10.307&     0.021&      0.45&       2.4&       7.1&     28&     28\\
     27&      7.68&     0.02&     2.39&     1.30&     0.22&   -18.945&     0.005&   -11.610&     0.014&      0.56&       7.4&      50.0&     24&     26\\
     28$^{*}$&      7.49&     0.03&     0.65&     0.33&     0.22&   -18.954&     0.012&   -12.280&     0.009&      0.33&       0.6&       1.7&    &    \\
     29&      7.40&     0.01&     1.09&     0.57&     0.03&   -18.938&     0.052&   -11.020&     0.028&      0.48&       2.1&       5.8&    &    \\
     30$^{*}$&      7.34&     0.03&     1.30&     0.65&     0.44&    -0.645&     0.006&     2.829&     0.011&      0.51&       3.4&      15.1&     33&     30\\
     31$^{*}$&      6.81&     0.03&     1.30&     0.65&     0.44&   -19.467&     0.007&   -11.494&     0.022&      0.55&       3.3&      15.0&     34&     31\\
     32&      6.66&     0.03&     1.96&     0.66&     0.10&    -6.440&     0.003&     5.700&     0.005&      0.44&      11.2&      52.9&     35&     32\\
     33$^{*}$&      6.63&     0.03&     0.65&     0.33&     0.22&    -6.361&     0.004&     5.129&     0.006&      0.20&       0.4&       1.1&    &    \\
     34&      6.43&     0.03&     1.09&     0.61&     0.19&    -0.854&     0.008&     2.780&     0.025&      0.58&       5.6&      17.0&     37&    \\
     35&      6.37&     0.02&     1.74&     0.70&     0.07&   -21.551&     0.003&    -7.459&     0.009&      0.45&       5.6&      24.8&     40&     33\\
     36&      6.23&     0.05&     0.87&     0.46&     0.21&    -6.116&     0.007&     6.167&     0.019&      0.50&       2.0&       6.6&    &    \\
     37&      6.21&     0.07&     3.04&     1.14&     0.25&     0.424&     0.002&     5.899&     0.006&      0.44&      19.1&     158.9&     41&     36\\
     38&      6.21&     0.02&     1.30&     0.79&     0.20&    -1.286&     0.006&     5.333&     0.017&      0.52&       7.6&      32.4&     42&     38\\
     39&      5.58&     0.04&     1.30&     0.26&     0.10&     3.492&     0.015&    -0.133&     0.025&      0.61&       7.2&      16.5&     45&    \\
     40&      5.54&     0.01&     3.26&     1.25&     0.05&    -6.976&     0.002&     5.043&     0.004&      0.45&      56.0&     381.9&     47&     40\\
     41&      5.53&     0.01&     1.09&     0.65&     0.13&     5.167&     0.008&    -1.416&     0.021&      0.47&       7.3&      21.6&    &    \\
     42&      5.32&     0.06&     1.96&     0.62&     0.16&     3.166&     0.006&     0.938&     0.017&      0.56&       7.5&      40.5&     50&     42\\
     43$^{*}$&      5.30&     0.03&     0.65&     0.33&     0.22&     5.521&     0.004&    -2.062&     0.036&      0.63&       4.2&       8.0&    &    \\
     44&      5.26&     0.03&     1.52&     0.73&     0.22&    -0.651&     0.008&     3.347&     0.019&      0.51&       4.3&      23.8&     53&     41\\
     45&      5.26&     0.04&     1.30&     0.53&     0.18&    -0.419&     0.019&     5.212&     0.025&      0.64&       5.8&      26.5&     49&     43\\
     46$^{*}$&      5.12&     0.03&     0.87&     0.43&     0.30&     6.324&     0.008&    -1.470&     0.019&      0.56&       3.3&       8.1&     51&    \\
     47$^{*}$&      5.11&     0.03&     0.65&     0.33&     0.22&     1.289&     0.006&    -0.340&     0.022&      0.60&       2.3&       5.7&    &    \\
     48&      5.11&     0.01&     1.09&     0.57&     0.07&     5.578&     0.008&     0.005&     0.014&      0.56&       3.4&      12.4&    &    \\
     49&      4.89&     0.03&     1.09&     0.42&     0.01&     1.665&     0.013&     0.062&     0.016&      0.62&      10.6&      28.4&     54&     51\\
     50$^{*}$&      4.83&     0.03&     0.87&     0.43&     0.30&     3.547&     0.019&     0.053&     0.033&      1.07&       8.2&      19.5&    &    \\
     51$^{*}$&      4.65&     0.03&     0.87&     0.43&     0.30&     2.120&     0.066&     0.800&     0.067&      0.68&       4.1&       7.7&    &    \\
     52&      4.65&     0.02&     1.52&     0.38&     0.05&     4.005&     0.008&     0.071&     0.015&      0.51&       9.0&      35.7&     55&     48\\
     53&      4.60&     0.02&     1.74&     0.56&     0.05&     6.033&     0.007&    -1.326&     0.011&      0.63&      21.9&      76.0&     59&     53\\
     54&      4.40&     0.01&     0.87&     0.27&     0.04&     1.777&     0.007&    -0.276&     0.042&      0.47&       2.1&       5.8&    &    \\
     55$^{*}$&      3.96&     0.03&     1.30&     0.65&     0.44&     5.789&     0.035&    -3.921&     0.054&      0.40&       1.3&       4.2&    &    \\
     56&      3.93&     0.05&     1.52&     1.12&     0.90&     1.432&     0.005&     0.359&     0.007&      0.49&      10.6&      46.0&    &    \\
     57&      3.90&     0.03&     1.52&     0.46&     0.09&     2.151&     0.005&     0.416&     0.021&      0.60&      15.0&      59.5&     65&     56\\
     58$^{*}$&      3.59&     0.03&     0.65&     0.33&     0.22&     5.700&     0.004&    -3.202&     0.038&      0.53&      14.2&      25.8&     62&    \\
     59&      3.52&     0.02&     1.74&     0.32&     0.06&     5.431&     0.004&    -1.619&     0.016&      0.55&      18.6&      74.0&    &    \\
     60$^{*}$&      3.51&     0.03&     1.09&     0.54&     0.37&     2.198&     0.008&    -0.150&     0.038&      0.59&       3.9&       6.0&    &    \\
     61$^{*}$&      3.49&     0.03&     0.87&     0.43&     0.30&     0.993&     0.015&    -0.072&     0.025&      0.58&       2.1&       3.8&     67&    \\
     62&      3.43&     0.03&     1.30&     0.66&     0.10&     1.825&     0.022&     0.472&     0.008&      0.62&      23.1&      74.7&     66&    \\
     63&      3.39&     0.04&     3.48&     1.76&     0.24&     5.914&     0.011&    -3.121&     0.007&      0.63&      56.1&     441.6&    &    \\
     64$^{*}$&      3.23&     0.03&     0.87&     0.43&     0.30&     0.547&     0.017&    -0.370&     0.024&      0.62&       3.5&       7.2&    &    \\
     65&      3.22&     0.01&     1.09&     0.50&     0.04&     0.380&     0.004&     5.093&     0.012&      0.47&       3.9&      13.4&     70&     60\\
     66&      2.97&     0.01&     1.09&     0.46&     0.03&     5.153&     0.007&    -2.083&     0.025&      0.56&      13.6&      37.8&    &    \\
     67&      2.93&     0.02&     1.30&     0.67&     0.12&   -11.555&     0.008&     1.372&     0.013&      0.53&       3.9&      17.6&     71&     61\\
     68&      2.86&     0.01&     1.52&     0.77&     0.06&     5.277&     0.005&    -3.978&     0.017&      0.52&       5.4&      31.5&     68&     64\\
     69&      2.84&     0.02&     1.30&     0.44&     0.07&     1.923&     0.020&     0.395&     0.017&      0.66&      12.2&      35.3&    &    \\
     70&      2.79&     0.03&     1.96&     0.91&     0.20&     5.183&     0.004&    -1.006&     0.007&      0.49&      11.0&      56.2&     73&    \\
     71$^{*}$&      2.75&     0.03&     1.30&     0.65&     0.44&     4.422&     0.009&    -9.763&     0.008&      0.54&       6.4&      29.1&     77&     66\\
     72&      2.63&     0.01&     1.09&     0.22&     0.01&    -0.220&     0.007&    -0.402&     0.016&      0.57&       4.3&      11.5&    &    \\
     73&      2.59&     0.04&     1.09&     0.65&     0.41&    -7.691&     0.004&     6.941&     0.012&      0.47&       2.6&      10.3&    &    \\
     74&      2.33&     0.01&     1.30&     1.55&     0.19&    -0.077&     0.007&     0.559&     0.020&      0.65&       4.2&      14.8&     82&    \\
     75&      2.32&     0.01&     1.09&     1.32&     1.79&    -9.318&     0.006&     5.233&     0.020&      0.59&       3.7&      14.0&     79&     71\\
     76&      2.26&     0.06&     2.17&     0.91&     0.25&     4.755&     0.003&    -2.500&     0.009&      0.50&      10.2&      51.5&     76&     62\\
     77&      2.08&     0.02&     1.74&     0.62&     0.06&    -3.385&     0.004&    -0.186&     0.004&      0.45&      16.7&      65.8&     81&     74\\
     78&      2.02&     0.03&     0.87&     0.26&     0.18&    -9.049&     0.008&     5.693&     0.027&      0.67&       3.0&       7.8&    &    \\
     79$^{*}$&      1.77&     0.03&     0.87&     0.43&     0.30&    -8.574&     0.005&     5.706&     0.018&      0.21&       0.6&       2.1&    &    \\
     80&      1.61&     0.03&     0.87&     0.33&     0.22&    -3.717&     0.020&     1.073&     0.037&      0.59&       2.9&       9.9&    &    \\
    81&      1.60&     0.04&     1.09&     0.51&     0.30&    -9.622&     0.004&     4.514&     0.012&      0.44&       1.9&       8.1&    &    \\
     82$^{*}$&      1.53&     0.03&     0.87&     0.43&     0.30&    -8.571&     0.017&     6.949&     0.024&      0.42&       2.7&       6.2&    &    \\
     83&      1.49&     0.02&     2.17&     0.89&     0.09&    -8.643&     0.002&     6.317&     0.004&      0.49&      26.6&     127.8&     84&     78\\
     84&      1.32&     0.02&     1.09&     1.61&     3.57&     4.748&     0.006&    -9.974&     0.008&      0.50&      10.9&      41.3&     85&     79\\
     85$^{*}$&      1.21&     0.03&     0.65&     0.33&     0.22&     0.591&     0.011&     0.287&     0.020&      0.52&       2.0&       5.8&    &    \\
     86&      1.20&     0.01&     0.87&     0.40&     0.04&     4.802&     0.008&    -5.389&     0.028&      0.52&       2.3&       7.9&    &    \\
     87$^{*}$&      1.17&     0.03&     0.65&     0.33&     0.22&     4.922&     0.010&    -9.204&     0.009&      0.21&       0.5&       1.4&    &    \\
     88&      1.04&     0.06&     1.30&     0.72&     0.35&     6.219&     0.005&   -10.002&     0.013&      0.50&       3.2&      12.2&     86&     84\\
     89&      0.96&     0.03&     1.09&     0.51&     0.12&     5.990&     0.016&   -10.074&     0.023&      0.48&       2.6&       9.4&    &    \\
     90$^{*}$&      0.76&     0.03&     0.65&     0.33&     0.22&    -7.961&     0.051&     7.123&     0.022&      0.58&       3.0&       8.2&    &    \\
     91&      0.65&     0.03&     2.82&     0.95&     0.09&     3.928&     0.004&    -5.107&     0.010&      0.50&      14.4&     120.9&     88&     82\\
     92$^{*}$&      0.61&     0.03&     0.65&     0.33&     0.22&    -8.511&     0.011&     6.655&     0.011&      0.50&       3.9&       7.9&    &    \\
     93&      0.58&     0.03&     1.96&     0.94&     0.17&     5.000&     0.004&    -9.749&     0.004&      0.52&      16.3&      87.5&     90&     88\\
     94$^{*}$&      0.53&     0.03&     1.09&     0.54&     0.37&     3.803&     0.006&    -4.631&     0.060&      0.67&       8.2&      10.9&     94&    \\
     95&      0.51&     0.06&     1.52&     1.92&     0.92&     3.502&     0.011&    -4.979&     0.017&      0.53&      10.7&      54.8&    &    \\
     96&      0.37&     0.14&     1.30&     0.74&     0.46&     4.263&     0.005&    -3.463&     0.018&      0.49&       3.4&      13.4&     98&     89\\
     97$^{*}$&      0.34&     0.03&     0.65&     0.33&     0.22&     0.004&     0.009&    -2.021&     0.024&      0.52&       1.8&       4.8&    &    \\
     98$^{r}$&     -0.01&     0.01&     3.04&     0.69&     0.01&     0.000&     0.001&     0.000&     0.003&      0.45&      66.7&     292.8&     96&     91\\
     99$^{*}$&     -0.12&     0.03&     0.87&     0.43&     0.30&     4.364&     0.017&    -3.247&     0.024&      0.54&       1.8&       4.8&    &    \\
    100$^{*}$&     -0.21&     0.03&     1.30&     0.65&     0.44&     0.068&     0.029&    -0.535&     0.012&      0.31&       2.3&       5.0&    &    \\
    101&     -0.74&     0.03&     0.87&     0.37&     0.08&    -8.951&     0.006&     3.190&     0.017&      0.49&       1.6&       5.2&    &    \\
    102&     -0.79&     0.03&     1.74&     1.07&     0.36&     5.102&     0.002&    -9.821&     0.006&      0.48&      10.7&      51.5&    103&     95\\
    103&     -1.23&     0.07&     2.17&     0.77&     0.21&    -8.289&     0.005&     1.792&     0.011&      0.52&       4.3&      26.2&    104&     98\\
    104&     -1.26&     0.02&     1.30&     1.00&     0.54&    -7.949&     0.006&     2.980&     0.015&      0.48&       2.0&       9.5&    &    \\
    105$^{*}$&     -1.85&     0.03&     0.65&     0.33&     0.22&     1.513&     0.005&   -17.936&     0.014&      0.46&       1.3&       3.3&    &    \\
    106&     -1.87&     0.02&     2.17&     0.84&     0.07&     6.518&     0.004&    -8.514&     0.007&      0.45&       6.3&      34.8&    106&    100\\
    107&     -2.31&     0.03&     1.09&     0.48&     0.21&     1.943&     0.003&   -18.748&     0.008&      0.44&       2.7&       9.4&    105&    102\\
    108&     -3.76&     0.05&     2.17&     1.21&     0.30&     2.643&     0.006&    -4.849&     0.007&      0.50&       7.6&      50.4&    109&    103\\
    109&     -5.99&     0.02&     1.96&     0.66&     0.04&   -14.404&     0.004&     5.277&     0.014&      0.48&       3.1&      17.0&    110&    104\\
    110&     -6.92&     0.01&     1.09&     0.46&     0.12&   -14.811&     0.005&     5.100&     0.013&      0.46&       1.3&       7.2&    &    \\
\enddata

\tablecomments{The representations of columns (1)-(14) are the
same as Table 2.}

\tablenotetext{*}{Feature which can not be well represented by a
Gaussian curve.}

\tablenotetext{r} {Reference feature.}
\end{deluxetable}

\begin{deluxetable}{lrrrrrrrrrrrrrrr}

\tablewidth{0pt} \tablecaption{\small 43 GHz SiO maser features
around VX Sgr observed by VLBA on 1999 May 31.}

\tablehead{ID&\multicolumn{1}{r}{V$_{LSR}$}& $\sigma_{V_{LSR}}$
&$\Delta u$&$\Delta$V& $\sigma_{\Delta
V}$ &\multicolumn{1}{c}{x}&$\sigma_{x}$ &\multicolumn{1}{c}{y}&$\sigma_{y}$&\multicolumn{1}{c}{$L$}&P&S&\multicolumn{2}{c}{Match ID} \\
\cline {14-15}

&\multicolumn{2}{c}{(km s$^{-1}$)}&(km
s$^{-1}$)&\multicolumn{2}{c}{(km s$^{-1}$)}& \multicolumn{2}{c}{(mas)}&\multicolumn{2}{c}{(mas)}&(mas)&(Jy)&(Jy km s$^{-1}$)&Epoch 1&2\\
(1)&(2)&(3)&(4)&(5)&(6)&\multicolumn{1}{c}{(7)}&(8)&\multicolumn{1}{c}{(9)}&(10)&(11)&(12)&(13)&\multicolumn{2}{c}{(14)}
}\startdata

      1$^{*}$&     14.34&     0.03&     0.87&     0.43&     0.30&    -1.155&     0.006&   -14.441&     0.024&      0.71&       4.5&      15.0&    &    \\
      2$^{*}$&     14.33&     0.03&     0.87&     0.43&     0.30&    -0.799&     0.009&   -18.578&     0.025&      0.53&       1.6&       5.6&    &    \\
      3$^{*}$&     13.36&     0.03&     0.65&     0.33&     0.22&    -1.081&     0.009&   -13.426&     0.029&      0.34&       0.9&       2.3&    &    \\
      4&     12.96&     0.02&     1.96&     1.28&     0.14&    -1.271&     0.003&   -14.575&     0.010&      0.46&       6.5&      40.2&    &    \\
      5&     12.86&     0.03&     2.61&     1.64&     0.39&    -1.027&     0.003&   -14.067&     0.009&      0.48&       8.2&      60.5&     11&      8\\
      6$^{*}$&     12.64&     0.03&     0.65&     0.33&     0.22&    -2.708&     0.031&     0.607&     0.015&      0.54&       4.3&      12.4&    &    \\
      7&     11.24&     0.02&     2.17&     0.70&     0.06&    -2.494&     0.011&     1.141&     0.019&      0.51&      11.4&      78.7&     13&     11\\
      8$^{*}$&     11.19&     0.03&     0.87&     0.43&     0.30&    -2.481&     0.013&     1.679&     0.009&      0.15&       0.8&       2.3&    &    \\
      9&     10.91&     0.03&     1.09&     0.45&     0.15&    -2.832&     0.008&     1.797&     0.028&      0.47&       8.1&      25.2&    &    \\
     10$^{*}$&     10.73&     0.03&     0.65&     0.33&     0.22&   -17.251&     0.006&   -13.271&     0.023&      0.51&       1.8&       4.8&     15&     12\\
     11&     10.60&     0.03&     1.09&     0.43&     0.11&     2.732&     0.024&    -6.018&     0.026&      0.52&       2.2&       8.8&    &    \\
     12$^{*}$&     10.02&     0.03&     1.09&     0.54&     0.37&   -18.617&     0.013&   -10.822&     0.030&      0.56&       2.8&       9.1&    &    \\
     13&      9.94&     0.02&     1.52&     0.45&     0.05&    -2.408&     0.007&     1.695&     0.015&      0.48&      13.2&      38.7&    &    \\
     14&      9.65&     0.01&     1.52&     0.74&     0.06&    -2.500&     0.014&     2.480&     0.018&      0.53&      16.4&      75.4&     19&     16\\
     15&      9.58&     0.03&     0.87&     0.63&     0.43&    -2.434&     0.010&     1.846&     0.021&      0.22&       2.1&       3.9&    &    \\
     16&      9.33&     0.03&     0.87&     0.13&     0.09&   -18.876&     0.009&    -9.876&     0.023&      0.60&       3.0&      10.8&    &    \\
     17&      9.20&     0.01&     1.09&     0.69&     0.10&    -1.998&     0.016&     2.068&     0.029&      0.49&      18.0&      56.8&    &    \\
     18&      9.11&     0.01&     1.30&     0.60&     0.06&    -0.848&     0.002&     3.495&     0.010&      0.49&      14.4&      49.8&     21&     20\\
     19&      9.08&     0.02&     0.87&     0.26&     0.07&    -0.898&     0.006&     4.148&     0.009&      0.31&       1.1&       3.2&    &    \\
     20$^{*}$&      8.88&     0.03&     0.87&     0.43&     0.30&   -18.756&     0.006&   -11.103&     0.022&      0.62&       4.3&      10.8&    &    \\
     21&      8.35&     0.02&     1.96&     0.88&     0.11&    -1.771&     0.008&     2.488&     0.018&      0.55&      13.3&      61.0&    &    \\
     22&      8.35&     0.01&     1.74&     0.56&     0.04&    -1.557&     0.013&     3.762&     0.019&      1.11&      12.1&      42.8&    &    \\
     23$^{*}$&      8.33&     0.03&     1.09&     0.54&     0.37&    -1.920&     0.009&     3.195&     0.023&      0.30&       1.7&       5.9&    &    \\
     24&      8.20&     0.03&     1.09&     0.48&     0.32&   -19.002&     0.024&   -12.257&     0.023&      0.32&       0.6&       2.2&    &    \\
     25&      8.05&     0.02&     2.82&     1.15&     0.10&    -1.969&     0.006&     3.797&     0.011&      0.50&      11.3&      74.8&     29&     25\\
     26&      7.96&     0.06&     3.26&     1.24&     0.26&   -18.931&     0.005&   -11.600&     0.015&      0.50&       6.2&      51.6&     24&     27\\
     27&      7.92&     0.01&     1.52&     0.65&     0.05&   -18.575&     0.004&   -10.408&     0.009&      0.43&       3.2&      20.0&    &    \\
     28&      7.74&     0.03&     1.09&     1.28&     4.72&   -19.056&     0.005&   -10.322&     0.020&      0.48&       2.8&      10.8&     28&     26\\
     29$^{*}$&      7.29&     0.03&     0.65&     0.33&     0.22&   -19.071&     0.016&   -10.997&     0.030&      0.42&       1.6&       3.3&    &    \\
     30&      6.92&     0.06&     1.96&     0.45&     0.17&    -0.712&     0.014&     2.683&     0.033&      0.57&       4.6&      14.4&     33&     30\\
     31&      6.68&     0.02&     1.52&     0.29&     0.06&   -19.418&     0.010&   -11.416&     0.036&      0.54&       4.4&      15.9&     34&     31\\
     32&      6.65&     0.01&     1.74&     0.73&     0.06&    -6.451&     0.003&     5.673&     0.006&      0.41&       7.8&      34.8&     35&     32\\
     33&      6.32&     0.01&     1.52&     0.67&     0.04&   -21.521&     0.002&    -7.460&     0.008&      0.42&       6.2&      27.4&     40&     35\\
     34&      6.31&     0.11&     2.17&     0.55&     0.25&    -0.782&     0.016&     2.936&     0.024&      0.58&       5.0&      18.7&    &    \\
     35$^{*}$&      6.20&     0.03&     0.65&     0.33&     0.22&     0.006&     0.009&     5.287&     0.028&      0.45&       1.0&       2.9&    &    \\
     36&      6.13&     0.04&     3.04&     1.27&     0.19&     0.411&     0.002&     5.883&     0.005&      0.41&      23.9&     173.4&     41&     37\\
     37&      6.11&     0.04&     1.52&     0.37&     0.11&    -6.156&     0.005&     6.260&     0.015&      0.48&       3.7&      17.3&    &    \\
     38&      5.94&     0.02&     1.52&     0.49&     0.07&    -1.264&     0.006&     5.367&     0.017&      0.50&       7.0&      26.6&     42&     38\\
     39$^{*}$&      5.61&     0.03&     0.87&     0.43&     0.30&     5.160&     0.003&    -1.382&     0.016&      0.48&       5.1&      16.3&    &    \\
     40&      5.45&     0.04&     3.04&     1.20&     0.15&    -6.982&     0.001&     5.029&     0.004&      0.43&      44.9&     310.3&     47&     40\\
     41&      5.34&     0.03&     1.09&     1.12&     3.50&    -0.667&     0.004&     3.358&     0.018&      0.47&       3.7&      13.2&     53&     44\\
     42&      5.22&     0.03&     1.96&     0.95&     0.11&     3.189&     0.008&     0.917&     0.018&      0.56&       5.3&      35.0&     50&     42\\
     43&      5.21&     0.01&     1.52&     0.86&     0.08&    -0.417&     0.012&     5.166&     0.019&      0.53&       6.2&      29.2&     49&     45\\
     44$^{*}$&      5.10&     0.03&     1.09&     0.54&     0.37&     0.457&     0.005&     6.084&     0.015&      0.51&       9.4&      25.9&    &    \\
     45&      5.07&     0.01&     1.09&     1.07&     0.92&     5.549&     0.008&     0.018&     0.012&      0.49&       3.2&      12.9&    &    \\
     46&      5.06&     0.10&     0.87&     0.27&     0.24&     5.541&     0.008&    -2.314&     0.036&      1.03&      13.8&      33.0&    &    \\
     47$^{*}$&      5.06&     0.03&     0.65&     0.33&     0.22&     5.214&     0.047&    -1.518&     0.011&      0.49&       3.2&       5.6&    &    \\
     48&      4.90&     0.02&     1.09&     1.36&     3.58&     3.794&     0.023&    -0.012&     0.041&      0.66&       3.3&      12.7&     55&     52\\
     49&      4.88&     0.01&     1.52&     1.72&     1.32&     6.223&     0.007&    -1.330&     0.017&      0.46&       4.8&      19.4&    &    \\
     50$^{*}$&      4.81&     0.03&     1.09&     0.54&     0.37&     3.449&     0.021&    -0.003&     0.034&      0.55&       3.9&      10.8&    &    \\
     51&      4.80&     0.11&     1.09&     0.41&     0.41&     1.791&     0.008&    -0.127&     0.027&      0.50&       2.0&       6.2&     54&     49\\
     52&      4.74&     0.03&     1.52&     0.74&     0.22&     4.011&     0.006&     0.029&     0.012&      0.44&       5.4&      27.9&    &    \\
     53&      4.57&     0.01&     1.52&     0.72&     0.05&     5.984&     0.005&    -1.325&     0.010&      0.46&      11.6&      42.6&     59&     53\\
     54&      4.22&     0.03&     1.74&     0.54&     0.08&     1.446&     0.012&     0.308&     0.014&      0.48&      13.8&      65.0&    &    \\
     55$^{*}$&      3.91&     0.03&     1.09&     0.54&     0.37&     5.525&     0.007&    -1.481&     0.015&      0.68&      17.2&      45.4&    &    \\
     56&      3.89&     0.06&     2.39&     1.58&     0.85&     2.153&     0.009&     0.544&     0.013&      0.51&      11.9&      66.0&     65&     57\\
     57&      3.56&     0.03&     0.87&     0.39&     0.26&     5.553&     0.015&    -3.821&     0.014&      0.60&       4.5&      12.4&    &    \\
     58&      3.55&     0.03&     3.69&     1.84&     0.18&     5.947&     0.008&    -3.132&     0.004&      0.53&      51.6&     410.2&    &    \\
     59$^{*}$&      3.47&     0.03&     0.87&     0.43&     0.30&     2.210&     0.012&    -0.227&     0.018&      0.34&       1.3&       3.8&    &    \\
     60&      3.14&     0.01&     0.87&     0.29&     0.01&     0.378&     0.004&     5.068&     0.014&      0.44&       3.1&       7.9&     70&     65\\
     61&      3.01&     0.01&     1.30&     1.42&     0.93&   -11.580&     0.007&     1.435&     0.016&      0.50&       3.0&      13.9&     71&     67\\
     62$^{*}$&      3.00&     0.03&     0.87&     0.43&     0.30&     4.792&     0.025&    -2.460&     0.042&      0.51&       3.3&       8.6&     76&     76\\
     63&      2.78&     0.01&     1.96&     1.18&     0.13&     5.125&     0.005&    -2.129&     0.017&      0.49&       6.6&      27.3&    &    \\
     64&      2.77&     0.05&     1.30&     2.03&     1.21&     5.271&     0.005&    -4.017&     0.019&      0.49&       3.0&      13.6&     68&     68\\
     65&      2.75&     0.02&     1.09&     0.42&     0.09&    -0.263&     0.005&    -0.451&     0.013&      0.45&       3.1&      10.9&    &    \\
     66&      2.71&     0.06&     1.30&     1.10&     0.85&     4.368&     0.005&    -9.760&     0.008&      0.53&       6.5&      25.8&     77&     71\\
     67&      2.69&     0.02&     1.09&     0.89&     1.18&     5.125&     0.005&    -0.451&     0.008&      0.21&       0.6&       2.2&    &    \\
     68&      2.68&     0.02&     1.30&     1.94&     3.08&     0.297&     0.009&     0.884&     0.017&      0.55&       2.7&      12.8&    &    \\
     69&      2.42&     0.07&     1.52&     0.50&     0.20&    -7.687&     0.005&     6.960&     0.014&      0.44&       2.8&      15.4&    &    \\
     70$^{*}$&      2.36&     0.03&     1.09&     0.54&     0.37&     5.118&     0.011&    -2.799&     0.038&      0.53&       1.4&       3.9&    &    \\
     71$^{*}$&      2.31&     0.03&     0.65&     0.33&     0.22&    -9.334&     0.015&     5.246&     0.052&      0.74&       3.3&       8.0&     79&     75\\
     72&      2.09&     0.02&     1.96&     0.77&     0.10&     4.746&     0.003&    -2.487&     0.013&      0.51&      10.6&      56.3&    &    \\
     73&      2.09&     0.03&     1.30&     1.90&     4.34&    -0.111&     0.007&     0.554&     0.015&      0.52&       4.0&      19.2&    &    \\
     74&      2.03&     0.01&     1.52&     0.62&     0.06&    -3.393&     0.004&    -0.167&     0.005&      0.41&      13.5&      49.5&     81&     77\\
     75$^{*}$&      1.99&     0.03&     0.87&     0.43&     0.30&    -3.553&     0.019&     0.563&     0.047&      0.35&       2.0&       5.6&    &    \\
     76&      1.83&     0.06&     1.30&     0.60&     0.31&    -9.629&     0.005&     4.477&     0.014&      0.44&       1.8&       8.7&    &    \\
     77&      1.75&     0.03&     1.08&     0.37&     0.11&     4.725&     0.010&    -1.812&     0.017&      0.30&       1.5&       5.5&    &    \\
     78&      1.40&     0.01&     1.95&     0.95&     0.11&    -8.638&     0.003&     6.337&     0.004&      0.48&      26.0&     122.6&     84&     83\\
     79&      1.39&     0.04&     1.30&     1.98&     6.29&     4.773&     0.019&    -9.910&     0.013&      0.49&       9.8&      45.8&     85&     84\\
     80&      1.31&     0.03&     1.52&     2.00&     3.01&     4.796&     0.007&    -5.491&     0.026&      0.51&       1.8&       9.8&    &    \\
     81$^{*}$&      1.21&     0.03&     0.65&     0.33&     0.22&    -3.746&     0.018&     1.087&     0.022&      0.57&       2.5&       5.9&    &    \\
     82&      1.14&     0.05&     1.52&     0.39&     0.14&     3.849&     0.005&    -4.973&     0.007&      0.49&      10.2&      51.1&     88&     91\\
     83$^{*}$&      0.98&     0.03&     0.65&     0.33&     0.22&     3.755&     0.025&    -4.222&     0.056&      0.44&       2.0&       4.6&    &    \\
     84&      0.79&     0.03&     1.74&     0.80&     0.19&     6.187&     0.005&   -10.030&     0.013&      0.54&       4.5&      21.0&     86&     88\\
     85&      0.74&     0.01&     1.09&     0.49&     0.02&     5.954&     0.021&   -10.164&     0.024&      0.48&       1.2&       3.5&    &    \\
     86&      0.70&     0.02&     1.09&     0.67&     0.23&     3.420&     0.006&    -5.000&     0.015&      0.51&       3.2&      13.0&    &    \\
     87&      0.53&     0.03&     1.09&     1.44&     1.72&    -7.879&     0.011&     7.185&     0.017&      0.57&       4.6&      14.2&    &    \\
     88&      0.48&     0.02&     2.17&     0.91&     0.07&     5.020&     0.005&    -9.761&     0.006&      0.52&      14.2&      64.7&     90&     93\\
     89&      0.19&     0.02&     2.17&     1.84&     0.86&     4.239&     0.007&    -3.375&     0.012&      0.51&       4.8&      32.1&     98&     96\\
     90$^{*}$&     -0.04&     0.03&     0.87&     0.43&     0.30&     3.652&     0.011&    -5.212&     0.034&      0.71&       6.9&      22.5&    &    \\
     91$^r$&     -0.07&     0.01&     2.61&     0.67&     0.01&     0.000&     0.001&     0.000&     0.004&      0.43&      51.7&     219.6&     96&     98\\
     92&     -0.16&     0.01&     2.61&     1.83&     0.29&     3.900&     0.006&    -5.191&     0.014&      0.50&      13.5&      95.1&    &    \\
     93$^{*}$&     -0.52&     0.03&     1.09&     0.54&     0.37&     5.026&     0.010&    -9.175&     0.012&      0.26&       0.8&       3.4&    &    \\
     94$^{*}$&     -0.56&     0.03&     0.65&     0.33&     0.22&   -10.277&     0.007&     2.154&     0.017&      0.44&       1.2&       3.3&    &    \\
     95&     -0.77&     0.03&     1.74&     0.87&     0.17&     5.071&     0.007&    -9.820&     0.005&      0.46&       9.7&      46.6&    103&    102\\
     96$^{*}$&     -0.80&     0.03&     0.65&     0.33&     0.22&     5.165&     0.009&   -10.461&     0.020&      0.23&       0.5&       1.2&    &    \\
     97&     -0.92&     0.01&     1.08&     0.31&     0.01&     7.218&     0.008&   -10.699&     0.018&      0.45&       2.1&       7.4&    &    \\
     98&     -1.17&     0.03&     1.74&     0.75&     0.15&    -8.310&     0.005&     1.840&     0.016&      0.53&       2.9&      17.4&    104&    103\\
     99&     -1.30&     0.03&     0.87&     0.14&     0.09&    -7.956&     0.004&     2.987&     0.010&      0.45&       3.0&       6.9&    &    \\
    100&     -1.85&     0.02&     1.96&     0.85&     0.10&     6.521&     0.003&    -8.519&     0.006&      0.42&       6.1&      32.8&    106&    106\\
    101&     -1.95&     0.02&     0.87&     1.05&     2.03&     1.521&     0.007&   -17.894&     0.015&      0.44&       1.2&       4.3&    &    \\
    102&     -2.32&     0.01&     1.09&     1.27&     0.22&     1.936&     0.004&   -18.712&     0.010&      0.41&       1.9&       7.5&    105&    107\\
    103&     -3.69&     0.04&     1.96&     1.11&     0.33&     2.599&     0.006&    -4.839&     0.006&      0.45&       8.0&      48.7&    109&    108\\
    104&     -6.03&     0.01&     1.30&     0.59&     0.06&   -14.399&     0.005&     5.342&     0.011&      0.46&       2.9&      12.7&    110&    109\\
    105&     -6.89&     0.03&     1.30&     0.80&     0.42&   -14.811&     0.004&     5.139&     0.010&      0.42&       1.9&       9.0&    &    \\

\enddata
\tablecomments{The representations of columns (1)-(14) are the
same as Table 2.}
 \tablenotetext{*}{Feature which can not be well
represented by a Gaussian curve.}

\tablenotetext{r} {Reference feature.}

\end{deluxetable}


\begin{thebibliography}{}

\bibitem{bo} Boboltz, D. A., Diamond, P. J., \& Kemball, A. J.
1997, ApJ, 487, L147
\bibitem{bo} Boboltz, D. A., \& Wittkowski, M. 2005, ApJ, 618, 953
\bibitem{Bow} Bowen, G. H. 1988, ApJ, 329, 299
\bibitem{Ch} Chapman, J. M., \& Cohen, R. J. 1986, MNRAS, 220, 513
\bibitem{di94} Diamond, P. J., Kemball, A. J., Junor, W., Zensus, A. Benson, J., \& Dhawan, V. 1994, ApJ, 430,
L61
\bibitem{di} Diamond, P. J., \& Kemball, A. J. 2003, ApJ, 599,
1372
\bibitem{do} Doel, R. C., Gray M. D., Humphreys, E. M. L.,
Braithwaite, M. F., \& Field, D. 1995, A\&A, 302, 797
\bibitem{do} Doeleman, S. S., Lonsdale, C. J., \& Greenhill, L. J.
1998, ApJ, 494, 400
\bibitem{El} Elitzur, M., 1992, Astronomical Masers (Dordrecht:
Kluwer)

\bibitem{go} Goldreich, P., \& Keeley, D. A., 1972, ApJ, 174, 517
\bibitem{go} Goldreich, P., Kwan, J., 1974, ApJ, 190, 27
\bibitem {gr} Gray, M. D., Humphreys, E.
M. L., \& Field, D. 1995, Ap\&SS, 224, 63
\bibitem {gr}Gray, M. D., \& Humphreys, E. M. L. 2000, New Astron.,
5, 155
\bibitem{Gr} Greenhill, L. J., Colomer, F., Moran, J. M., Danchi, W. C., \& Bester, M. 1995, ApJ, 449, 365
\bibitem{Gw} Gwinn, C. R. 1994, ApJ, 429, 241
\bibitem{Hu} Humphreys, E. M. L., Gray, M. D., Yates, J. A., Field, D., Bowen, G., \& Diamond, P. J. 1996, MNRAS, 282, 1359
\bibitem{hu} Humphreys, E. M. L., Gray, M. D., Yates, J. A., Field,
D., Bowen, G. H., \& Diamond, P. J. 2002, A\&A, 386, 256

\bibitem{Im} Imai, H., Deguchi, S., \& Sasao, T. 2002, ApJ, 567,
971
\bibitem{ka} Kamohara, R., Deguchi, S., Miyoshi, M., \& Shen, Z.-Q., 2005, PASJ,
57, 341
\bibitem{ku} Kukarkin, B. V., Khopopov, P. N., Efremov, Yu, N., Kukarkina,
N.P., Kurochkin, N. E., Medvedeva, G. I., Perova, N. B.,
Federovich, V. P., \& Frolov, M. S. 1970. Catalogue of Variable
Stars, Astronomical Council of the Academy of Sciences in the
USSR, Moscow.
\bibitem{Mo} Monnier, J. D. et al. 2004, ApJ, 605, 436
\bibitem{Mu} Murakawa, K., Yates, J. A., Richards, A. M. S., \& Cohen, R .
J. 2003, MNRAS, 344, 1
\bibitem{ph} Phillips, R. B., Straughn, A. H., Doeleman, S. S., \&
Lonsdale, C. J. 2003, ApJ, 588, L105
\bibitem{re} Reid, M. J., \& Menten, K. M. 1997, ApJ, 476, 327
\bibitem{so} Soria-Ruiz, R., Alcolea, J., Colomer, F., Bujarrabal,
V., Desmurs, J. F., Marvel, K. B., \& Diamond, P. J. 2004, A\&A,
426, 131
\bibitem{W} Walker, R. C. 1984, ApJ, 280, 618


\end{thebibliography}
\end{document}